\documentclass{mn2e}
\usepackage{epsfig,amssymb,amsmath}

\newcommand{\beq}{\begin{eqnarray}}
\newcommand{\enq}{\end{eqnarray}}
\newcommand{\m}[1]{\bmath{#1}}
\newcommand{\pa}{\partial}
\newcommand{\bra}[1]{\left#1}
\newcommand{\ket}[1]{\vphantom{\sqrt{0}}\right#1}
\newcommand{\mtext}[1]{\hspace{0.6cm}\mbox{#1}\hspace{0.6cm}}
\newcommand{\hB}{\m{\hat{B}}}
\newcommand{\hn}{\m{\hat{n}}}
\newcommand{\hr}{\m{\hat{r}}}
\newcommand{\hx}{\m{\hat{x}}}
\newcommand{\hy}{\m{\hat{y}}}
\newcommand{\hz}{\m{\hat{z}}}
\newcommand{\hphi}{\m{\hat{\phi}}}

\begin{document}

\title{Toroidal magnetic fields in type II superconducting neutron stars}
\author[T. Akg\"{u}n and I. Wasserman]
{T.~Akg\"{u}n\thanks{E-mail: akgun@astro.cornell.edu; ira@astro.cornell.edu} and
I.~Wasserman\footnotemark[1]
\\Center for Radiophysics and Space Research, Cornell University, Ithaca, NY, 14853, USA}

\onecolumn

\maketitle

\begin{abstract}
We determine constraints on the form of axisymmetric toroidal magnetic fields dictated by
hydrostatic balance in a type II superconducting neutron star with a barotropic equation
of state. Using Lagrangian perturbation theory, we find the quadrupolar distortions due
to such fields for various models of neutron stars with type II superconducting and
normal regions. We find that the star becomes prolate and can be sufficiently distorted
to display precession with a period of the order of years. We also study the stability of
such fields using an energy principle, which allows us to extend the stability criteria
established by R.~J. Tayler for normal conductors to more general media with magnetic
free energy that depends on density and magnetic induction, such as type II
superconductors. We also derive the growth rate and instability conditions for a specific
instability of type II superconductors, first discussed by P. Muzikar, C.~J. Pethick and
P.~H. Roberts, using a local analysis based on perturbations around a uniform background.
\end{abstract}

\begin{keywords}
Stars: neutron -- Magnetic fields -- Magnetohydrodynamics -- Dense matter
\end{keywords}

\section{Introduction}
Timing residuals varying on timescales of order months to years have been detected in
several pulsars, most spectacularly in PSR B1828--11, where several cycles of nearly
periodic variation have been reported (Stairs, Lyne \& Shemar 2000; Stairs et al. 2003).
For PSR B1828--11, the precession period is $P_p\approx 500\,{\rm d}\approx 4.3\times
10^7 \,{\rm s}$ and the spin period is $P_\star\approx 0.405\,{\rm s}$; interpreting the
long term timing residuals as rigid body precession then implies a stellar distortion
$\epsilon\approx P_\star/P_p\approx 9.4\times 10^{-9}$. Precession affects arrival times
in two ways (Cordes 1993; Akg\"{u}n, Link \& Wasserman 2006): (i) Geometrical residuals
arise because the pulsar beam crosses the plane formed by the angular momentum of the
star and the line of sight to the observer at times that vary periodically over the
precession cycle. (ii) Variations in the angle between the spin and magnetic axes result
in a periodic variation of the pulsar spindown torque, causing pulse arrival times to
vary periodically as well. Precession models that combine these two effects describe the
data from PSR B1828--11 adequately (Jones \& Andersson 2001; Link \& Epstein 2001;
Akg\"{u}n et al. 2006).

Problems with these models remain, however. One is the observation by Shaham (1977, 1986)
that vortex line pinning can prevent long period precession, substituting instead
precession with very short periods (of order 10--100 spin periods, rather than $10^8$)
that damps out after perhaps $10^4$ cycles, contrary to observations (Sedrakian,
Wasserman \& Cordes 1999). Although Link \& Cutler (2002) showed that the precession
amplitude in PSR B1828--11 may be large enough to unpin all vortex lines in the
crystalline stellar crust, Link (2003) argued that the interaction of (magnetized) core
superfluid vortex lines with the flux tubes in type II superconducting regions would also
prevent long period precession. One way out is that the core neutrons are not superfluid,
an idea that gets some support from comparing theoretical models for cooling neutron
stars with observations (e.g. Yakovlev \& Pethick 2004, and references therein).

Even if vortex line pinning is not an issue, the required stellar distortion is
problematic. Although the rotational distortion of a fluid star is substantial,
$\epsilon_{\rm rot} \approx E_{\rm rot}/E_{\rm grav}\approx 7\times 10^{-8}R_{6}^3/
M_{1.4}^{}P_\star^2$ (for uniform density), where $R_\star = 10^6 R_{6}\,{\rm cm}$ and
$M_\star = 1.4M_{1.4} M_\odot$ are the radius and mass, and $P_\star$ is the spin period
in seconds, the bulge in a slowly rotating, self-gravitating fluid is always axisymmetric
about the angular momentum axis, and cannot result in precession. Only the solid crust of
a neutron star can support distortions that are fixed in the rotating frame of the star,
as are needed for precession. However, the crust of a neutron star is not very rigid: its
shear modulus is only about 0.01 times the crustal pressure. Consequently, $\epsilon \ll
\epsilon_{\rm rot}$ if the crustal distortion is ``relaxed'' at the current rotational
frequency of the star (Baym \& Pines 1971; Cutler, Ushomirsky \& Link 2003). For PSR
B1828--11, agreement between the observed and calculated precession frequencies would
require that the crustal deformation be relaxed at a rotation frequency of about 40 Hz,
compared with the present frequency of about 2.5 Hz (Cutler et al. 2003).

An alternative explanation for the precession frequency is that it is due to stellar
distortions resulting from magnetic stresses. The idea that a rotating, magnetic star
must precess goes back about fifty years (e.g. Spitzer 1958). If the magnetic field and
rotational axes are not lined up, then the moment of inertia of the star is the sum of
two axisymmetric contributions that are misaligned: the rotational distortion, estimated
above, and a magnetic distortion of order $\epsilon_{\rm mag}=E_{\rm mag}/E_{\rm grav}$.
In such a case, the star will precess about the magnetic axis with a frequency
proportional to the magnetic distortion (Mestel \& Takhar 1972; Mestel et al. 1981;
Nittmann \& Wood 1981).

For the typical inferred dipole magnetic fields of neutron stars, the magnetic
deformation is far too small, and the resulting precession period is far too long:
$\epsilon_{\rm mag} \sim 10^{-12}B_{12}^2R_{6}^4M_{1.4}^{-2}$ for a dipole magnetic field
strength $B = 10^{12}B_{12}\, \rm G$. However, substantial internal toroidal fields (e.g.
$B_{12}\sim 100$) could lead to large enough magnetic distortions to account for the
precession frequency of PSR B1828--11 (e.g. Ioka 2001; Cutler 2002).

Larger magnetic deformations could also result from type II superconductivity in the
neutron star's core for a given magnetic induction strength in the superconductor (e.g.
Jones 1975; Easson \& Pethick 1977; Cutler 2002; Wasserman 2003). In this paper, we shall
examine the distortions of a fluid neutron star induced by the enhanced magnetic stresses
associated with type II superconductivity. Here we focus on primarily toroidal fields,
partly because they are easier to treat, but also because they lead to \emph{prolate}
stellar distortions, which the data on PSR B1828--11 seem to favor at least weakly
(Wasserman 2003; Akg\"{u}n et al. 2006). We will include a weaker poloidal component that
can leak into the stellar magnetosphere, as is required for the pulsar to be active.
Differential rotation within a newborn neutron star most likely amplifies the toroidal
component of the field (Thompson \& Duncan 2001), but stable configurations will require
some poloidal field as well (e.g. Braithwaite \& Nordlund 2006). We have developed the
(more complicated) formalism needed to treat purely poloidal fields in a compressible
type II superconductor (Akg\"{u}n 2007), and will present those calculations elsewhere.

As a result of $^1S_0$ pairing via strong interactions, the protons in the interior of a
neutron star are expected to form a type II superconductor at baryon number densities
between $\sim 0.1-0.6 \, {\rm fm^{-3}}$ (e.g. Baym, Pethick \& Pines 1969; Baym \&
Pethick 1975; Easson \& Pethick 1977; Elgar{\o}y et al. 1996; Jones 2006; Baldo \&
Schulze 2007). Magnetic flux penetrates the superconducting region in the neutron star in
the form of quantized magnetic flux tubes. Typically, in a neutron star the critical
field is $H_{\rm c1} \sim 10^{15}\, \rm G$, and the magnetic induction is $B\sim
10^{12}\, {\rm G} \ll H_{\rm c1}$, so the magnetic field is $H \approx H_{\rm c1}$ and is
approximately a function of baryon density (e.g. Easson \& Pethick 1977).

In the neutron star crust, which exists at densities below $\sim 2 \times 10^{14} \, \rm
g/cm^3$ (Baym, Bethe \& Pethick 1971; Lorenz, Ravenhall \& Pethick 1993), protons are
bound in nuclei, and as a result, superconductivity is suppressed. Magnetic stresses in a
type II superconductor are $\sim H B/4\pi \approx H_{\rm c1}B/4\pi$, and consequently
will be about $H_{\rm c1}/B \sim 10^3$ times larger than those in a normal conductor with
the same $B$, which scale as $B^2/8\pi$ (Jones 1975; Easson \& Pethick 1977). Stresses of
this magnitude are capable of distorting the neutron star sufficiently to cause
precession of the star with a period of the order of a year (Cutler 2002; Wasserman
2003). However, we note that hydrostatic equilibrium requires approximate continuity of
$HB$ throughout the star, so the induction $B_n$ in the normal region is much larger than
the induction $B_s$ in the superconducting region: $B_n \propto (HB_s)^{1/2} \gg B_s$.
Configurations with large discontinuities in stress are unstable, so it is unrealistic to
embed a superconducting region with an anomalously large stress inside a star with
otherwise much smaller stress.

The magnetic force in a type II superconductor is inherently different than in a normal
conductor. The difference results from the fact that the magnetic free energy in a type
II superconductor depends both on the magnetic induction, $B$ (or equivalently, $u_{\rm
mag} = B^2/8\pi$) and on the proton number density, $n_p$. The proton number density is a
function of the baryon number density, and consequently can be expressed as a function of
total mass density, $\rho$. A good approximation is to take $n_p\propto\rho$ (Easson \&
Pethick 1977). On the other hand, in a normal conductor the magnetic free energy is a
function of magnetic induction alone.

The purpose of this paper is to determine magnetic field configurations in neutron stars
with type II superconductors, \emph{consistent with hydrostatic balance}, and assess
their stability. We assume that the magnetic deformations are small, which enables a
perturbative treatment. We neglect rotational deformations, slow fluid motions and
associated viscous effects, which can be included at a later stage (extending methods
laid out by Mestel \& Takhar 1972; Mestel et al. 1981; Nittmann \& Wood 1981). With these
solutions we can determine the magnetic distortion explicitly (cf. Cutler 2002, who
expressed the distortions in terms of averages over unspecified field configurations).

Assuming (cold nuclear) matter with a barotropic equation of state $p(\rho)$ imposes
significant constraints on the possible variation of the magnetic induction $B(r,\theta)
$ in the star. This is because Euler's equation of magnetohydrostatic balance requires
that the magnetic force per unit mass be a total gradient (a result well known for normal
magnetic equilibria; see e.g. Prendergast 1956; Monaghan 1965). The fact that $H\approx
H_{\rm c1}(\rho)$ is a function of $r$ alone to lowest order further restricts the range
of possible $B(r,\theta)$. With these constraints, we can evaluate the quadrupolar
deformation of the star in hydrostatic balance (as well as other multipoles, which are
uninteresting for precession). In practice, we only calculate these for the $\gamma = 2$
polytropic equation of state $p = \kappa\rho^2$, where $\kappa$ is a constant, but the
formalism can be applied to any $p(\rho)$. Moreover, although we only present examples
for which $H = H_{\rm c1}(\rho)$, our formalism applies to any magnetic free energy
$F(\rho,B)$, hence $H = 4\pi\pa F/\pa B = H(\rho,B)$.

Even with the restrictions imposed by hydrostatic balance in a barotropic fluid, and the
density dependence of $H$, many possible $B(r,\theta)$ are permitted, even when we trim
the set of solutions by obvious requirements such as regularity. Stability ought to weed
out even more possibilities. To examine this question, we use the energy principle that
has proved fruitful for normal magnetic substances (e.g. Bernstein et al. 1958; Tayler
1973), extended to superconductors in which the magnetic free energy (and consequently
$H$) has arbitrary dependencies on $\rho$ and $B$. (Roberts 1981 examined this problem
for $H\propto\rho$.) From this stability criterion, we show that the most pernicious
axisymmetric instability is the interchange instability (just as in normal conductors),
and we show how the list of candidate field configurations can be winnowed further by
requiring immunity against it. The interchange instability can be viewed as a magnetic
buoyancy mode. Our detailed treatment of perturbations is applied specifically to
one-component fluids. Buoyancy due to multi-fluid composition, which arises as a result
of the density dependence of the number density of charged particles in chemical
equilibrium, will introduce new modes (Reisenegger \& Goldreich 1992), and may change the
interchange instability conditions (Ferri\`{e}re, Zimmer \& Blanc 1999, 2001). We
postpone a complete consideration of these effects to a later paper, but in
\S\ref{sectioncriteria} we argue that stability constraints on the toroidal field shape
remain the same.

For non-axisymmetric perturbations, the character of the energy principle is markedly
different in the superconducting case. From it we find a specific stability criterion for
what we will refer to as the Muzikar--Pethick--Roberts (MPR) instability first discussed
by Muzikar \& Pethick (1981) and Roberts (1981), who showed that for sufficiently weak
magnetic induction $B \lesssim 10^{13}\, \rm G$, the density dependence of $H$ promotes
the formation of domains with and without magnetic flux. From a local stability analysis,
we show that this instability only acts for $m > 0$ (non-axisymmetric) modes and only on
very small scales perpendicular to the field, corresponding to wave numbers $\sim
10^{4}/R_\star$. We estimate the growth time of the instability on these scales to be of
order $10^3 \, \rm s$ for typical parameters, i.e. longer than typical Alfv\'{e}n wave
crossing times. Although this is a distinctive mode associated with type II
superconductors, the fact that it only acts on small length scales may cause it to be
suppressed by small viscous effects. Moreover, since the instability is local it is
likely to be present in a rotating star as well. Preliminary calculations suggest that
while the stability condition is altered by buoyancy, the unstable MPR mode persists and
has the same growth rate as in a one-component fluid.

In this treatment, we neglect rotation and internal fluid motions. Our primary goal is to
understand the effects of the density dependence of the magnetic free energy $F(\rho,B)$
on equilibrium and stability. This case has been previously treated by Roberts (1981),
who considered poloidal fields in a completely type II superconducting star of uniform
density and magnetic field $H \propto \rho$. Here we extend these considerations to
barotropic equations of state and magnetic fields of the form $H(\rho,B)$ in fluid stars
with type II superconducting shells. We will be concerned with toroidal magnetic fields
in this paper, deferring the detailed treatment of poloidal fields to future work. We
then calculate explicitly the extent of stellar deformation due to the magnetic field.
Spitzer (1958) and Mestel \& Takhar (1972) argued that, to lowest order, the rotational
and magnetic deformations can be calculated separately. Then, a misalignment in the
rotational and magnetic deformations leads to precession, as mentioned above.

In addition to the proton superconductor, there may be a commingled neutron superfluid in
the core of a neutron star. If so, the two superfluids are coupled via entrainment. One
consequence is that the vortices in the neutron superfluid acquire magnetic flux and
therefore couple to the magnetic flux tubes in the proton superconductor. This
interaction is expected to impede precession (Link 2003). The long-term periodicity
observed in PSR B1828--11 may require this interaction to be of limited scale, perhaps
implying that there is no commingling of the two fluids. Moreover, theoretical models for
cooling neutron stars suggest that there is no compelling observational evidence for core
neutron superfluid (Yakovlev \& Pethick 2004). Although gap calculations generally
support the existence of a $^1S_0$ crustal neutron superfluid and a core proton
superconductor, the theory is less certain about the $^3P_2$ core neutron superfluid.
(Elgar{\o}y et al. 1996; Baldo \& Schulze 2007). Here, we assume that there is no core
neutron superfluid overlapping with regions of proton superconductivity. This simplifies
the problem, as the behavior of a mixed superfluid-superconductor system can be very
complex (Glampedakis, Andersson \& Jones 2007). Moreover, for the reasons given above,
this may even be justified.

Here, we are primarily concerned with the equilibrium structure of the magnetic field.
Although we will also discuss the stability from an energy principle point of view, we
will not delve into the more comprehensive treatment of modes which should also include
rotation, internal velocity fields, multi-fluid components, and the elastic crust, as
well as dissipation, mutual friction and entrainment effects, which would arise in a
superfluid-superconductor mixture. In particular, dissipation is strongly dependent on
whether the neutrons are superfluid or not. Moreover, there will be friction on the
magnetic flux tubes which is especially important if they coexist with neutron vortices.
Stability of rotating stars is known to be affected by normal magnetic fields
(Glampedakis \& Andersson 2007), and we expect the same to be true in the presence of
superconductivity. Therefore, our work is only a first step towards a more complete
treatment of the neutron star interior, where we highlight features arising from the
density dependence of the magnetic free energy.

The outline of this paper is as follows: in \S\ref{sectiontypeII}, we discuss the
magnetic stress tensor and force in a type II superconductor. In \S\ref{sectiontoroidal},
we determine the form of the toroidal magnetic fields in the normal and superconducting
regions, consistent with the boundary conditions at the stellar surface and internal
boundaries. We then proceed with the calculation of the hydrostatic equilibrium in the
presence of such magnetic fields in various neutron star models with type II and normal
regions. We calculate the density and gravitational potential perturbations and determine
the moments of inertia of the perturbed star. In \S\ref{sectionstability}, we discuss the
stability of toroidal fields in the normal and superconducting cases. We show that the
interchange instability is the worst axisymmetric instability, and derive the MPR
instability conditions and relevant time and length scales from a local analysis. In
\S\ref{sectionpoloidal}, we discuss the possibility of adding a small poloidal component
to help stabilize the toroidal fields. We derive the form of this poloidal field that is
consistent with the requirements that the magnetic force be a gradient and that the
magnetic induction be divergenceless.

\section{Magnetic Force in a Type II Superconductor}\label{sectiontypeII}
The magnetic stress tensor in a type II superconductor is given as (Easson \& Pethick
1977),
        \beq
        \sigma_{ij} = \left[F - \rho\frac{\pa F}{\pa\rho}
        - B\frac{\pa F}{\pa B}\right]\delta_{ij} + \frac{H_iB_j}{4\pi} \ .
        \label{typeIIstress}
        \enq
The magnetic free energy $F(\rho,B)$ is a function of mass density, $\rho$ and magnetic
induction, $B$. In isotropic media the magnetic field $H_i$ and induction $B_i$ are
parallel, so that $\sigma_{ij} = \sigma_{ji}$. In general, the relation between the
magnetic field and induction is given through (Josephson 1966),
        \beq
        H = 4\pi \frac{\pa F}{\pa B} \ .
        \label{typeIIfree}
        \enq
In a normal conducting medium we have $H = B$, i.e. the magnetic field is independent of
density, and the free energy is equal to the magnetic energy $F = B^2/8\pi$. Thus, the
stress tensor in this case reduces to,
        \beq
        \sigma_{ij} = - \frac{B^2}{8\pi}\delta_{ij} + \frac{B_iB_j}{4\pi} \ .
        \label{stressnormal}
        \enq
On the other hand, the magnetic field in a strongly type II superconducting medium, such
as the proton superconductor in a neutron star, is $H \approx H_{\rm c1} \gg B$, and
depends most sensitively on the proton number density $n_p$ and the superconducting
energy gap $\Delta$ (Tinkham 1975; Easson \& Pethick 1977), which are functions of baryon
density $\rho$ (Elgar{\o}y et al. 1996; Baldo \& Schulze 2007); therefore, $H \approx
H(\rho)$ and $F \approx HB/4\pi$. In this case, the magnetic stress tensor reduces to,
        \beq
        \sigma_{ij} = - \rho\frac{\pa F}{\pa\rho}\delta_{ij} +
        \frac{H_iB_j}{4\pi} \ .
        \label{stresstypeII}
        \enq
The stress tensor used by Roberts (1981) is of this form, with $H \propto \rho$.

In general, the gradient of the free energy is given as,
        \beq
        \nabla_iF = \frac{\pa F}{\pa \rho}\nabla_i\rho +
        \frac{\pa F}{\pa B}\nabla_iB \ .
        \enq
From equation (\ref{typeIIfree}) it follows that,
        \beq
        B\nabla_i\frac{\pa F}{\pa B} = \frac{B_k\nabla_i H_k}{4\pi} \ .
        \enq
Making use of these relations as well as the fact that $\m\nabla\cdot\m{B} = 0$, the
magnetic force density can be calculated from equation (\ref{typeIIstress}) as,
        \beq
        f_i = \nabla_j\sigma_{ij} =
        - \rho\nabla_i\frac{\pa F}{\pa\rho}
        - B\nabla_i\frac{\pa F}{\pa B}
        + \frac{B_j\nabla_jH_i}{4\pi}
        = \frac{[(\m\nabla\times\m{H})\times\m{B}]_i}{4\pi}
        - \rho\nabla_i\frac{\pa F}{\pa \rho} \ .
        \label{typeIIforce}
        \enq
This is the form of the force in a type II superconductor. (In fact, it is true in any
magnetic medium where the free energy is a function of density and magnetic induction.)
This is inherently different from the force in a normal conducting medium, which can be
retrieved by setting $H = B$ and $F = B^2/8\pi$.

In hydrostatic balance,
        \beq
        \m\nabla p + \rho\m\nabla\phi = \m{f}_{\rm mag} \ ,
        \label{eulertypeII}
        \enq
where $p$ is pressure, $\rho$ is mass density, $\phi$ is gravitational potential, and
$\m{f}_{\rm mag}$ is the magnetic force density (equation \ref{typeIIforce}). In
barotropic equations of state, pressure is a function of density and we can define $d h
(\rho) = \rho^{-1} d p (\rho)$; then,
        \beq
        \rho\m\nabla (h + \phi) = \m{f}_{\rm mag} \ .
        \label{eulerh}
        \enq
This equation requires the magnetic force per unit mass to be a gradient of a potential,
i.e. $\m{f}_{\rm mag} = - \rho\m\nabla\psi$. We will express the magnetic potential as
the sum of two terms,
        \beq
        \psi = \psi_{\rm I} + \psi_{\rm II} \ ,
        \label{pottypeII}
        \enq
where, we define,
        \beq
        \frac{(\m\nabla\times\m{H})\times\m{B}}{4\pi} =
        \frac{\m{J}\times\m{B}}{c} = - \rho\m\nabla\psi_{\rm I}
        \mtext{and}
        \psi_{\rm II} = \frac{\pa F}{\pa\rho} \ .
        \label{pottypeII2}
        \enq
$\m{J}$ is the current density, $\psi_{\rm I}$ is the magnetic potential for a normal
conductor, and $\psi_{\rm II}$ is present only for a type II superconductor. The second
term in the magnetic force (equation \ref{typeIIforce}) is already a gradient. On the
other hand, note that the requirement for the first term to be a gradient can be
expressed alternatively as,
        \beq
        \m\nabla\times\left(\frac{\m{J}\times\m{B}}{\rho c}\right) = 0 \ .
        \label{gradient}
        \enq
This equation needs to be satisfied for both the normal and type II superconducting
cases, and imposes a severe restriction on the form of the magnetic fields, which are
also required to satisfy $\m\nabla\cdot\m{B} = 0$. The normal conducting case is
discussed, for example, in Prendergast (1956) and Monaghan (1965). For the strongly type
II case and $H\propto\rho$, Roberts (1981) found poloidal field configurations for
uniformly dense stars, and Akg\"{u}n (2007) found poloidal field configurations for
$\gamma = 2$ polytropes.

\section{Toroidal Fields}\label{sectiontoroidal}
The current density for a toroidal field $\m{H} = H(r,\theta)\hphi$ is,
        \beq
        \frac{4\pi\m{J}}{c} = \m\nabla\times\m{H}
        = \m\nabla(Hr\sin\theta)\times\frac{\hphi}{r\sin\theta} \ .
        \label{toroidal-current}
        \enq
Taking the induction to be $\m{B} = B(r,\theta)\hphi$, we get,
        \beq
        \frac{\m{J}\times\m{B}}{\rho c} =
        \frac{(\m\nabla\times\m{H})\times\m{B}}{4\pi\rho}
        = - \frac{B\m\nabla(Hr\sin\theta)}{4\pi\rho r\sin\theta} \ .
        \label{toroidal-force}
        \enq
This is clearly a total gradient, as required by equation (\ref{gradient}), for magnetic
inductions of the form,
        \beq
        B(r,\theta) = 4\pi\rho r\sin\theta f(Hr\sin\theta) \ ,
        \label{toroidal-typeII}
        \enq
where $f$ is an arbitrary function of $\zeta = Hr\sin\theta$. The factor of $4\pi$ is
included so that defining a new function through $f(\zeta) = g'(\zeta)$ gives, using the
definitions in equation (\ref{pottypeII2}),
        \beq
        \frac{\m{J}\times\m{B}}{\rho c} = - \m\nabla g(\zeta)
        \mtext{i.e.} \psi_{\rm I}(r,\theta) = g(\zeta) \ .
        \label{psiI}
        \enq
This is valid for any $H(r,\theta)$. However, for a strongly type II superconductor $H
\approx H(r)$, and we have (equation \ref{pottypeII2}),
        \beq
        \psi_{\rm II} = \frac{B}{4\pi}\frac{d H}{d\rho}
        = \frac{d\ln H}{d\ln\rho}\zeta g'(\zeta) \ .
        \label{psiII}
        \enq

For a normal conductor $H = B$, and equation (\ref{toroidal-typeII}) implies that the
magnetic induction is now given through the form,
        \beq
        B(r,\theta) = \frac{h(\rho r^2\sin^2\theta)}{r\sin\theta} \ ,
        \label{toroidal-normal}
        \enq
where $h$ is an arbitrary function of $\xi = \rho r^2\sin^2\theta$. It then follows that,
        \beq
        \frac{\m{J}\times\m{B}}{\rho c} = - \m\nabla\psi =
        - \frac{\m\nabla h^2(\xi)}{8\pi\xi}
        \mtext{i.e.}
        \psi'(\xi) = \frac{h(\xi)h'(\xi)}{4\pi\xi} \ .
        \label{toroidal-normal2}
        \enq
Note that, for a uniform density, the magnetic induction is a function of the cylindrical
radius, $\varpi = r\sin\theta$.

\subsection{Star with a Superconducting Shell}

        \begin{figure}
        \centerline{\includegraphics[scale=0.6]{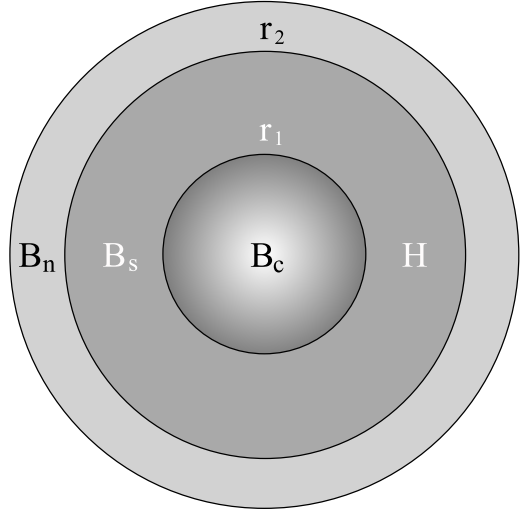}}
        \caption[A star with a normal core, superconducting shell, and a surrounding normal
        layer.]{A star with a normal core, superconducting shell, and a surrounding normal
        layer. The radius of the core is $r_1$ and the outer radius of the superconducting
        shell is $r_2$.}
        \label{sketch}
        \end{figure}

Consider the case of a strongly type II superconducting region confined to a spherical
shell between radii $r_1$ and $r_2$ (where $r_2 > r_1$). Let the magnetic field be $B_c$
inside the normal core, $H$ inside the superconducting shell (with a corresponding
magnetic induction $B_s$), and $B_n$ inside the normal outer layer (as depicted in
fig.~\ref{sketch}). Since the fields have no radial components in this case, they need
not be continuous across the boundaries, and there will be surface currents.

In fact, it turns out that in the toroidal case it is not possible to have a continuous
magnetic field across the boundaries, if $H = H(\rho)$ in the superconducting region.
Consider one of the boundaries of the superconducting shell, located at $r = r_b$. For
the present discussion, it is immaterial whether the normal region lies on the inside or
the outside of the boundary. In the absence of surface currents, the boundary condition
that follows from Maxwell's equations requires the continuity of the tangential magnetic
field,
        \beq
        \hr\times\m{H} = \hr\times\m{B}_n \ .
        \label{maxwell}
        \enq
Since $H$ is a function of radius in a strongly type II superconductor, for this equation
to be satisfied everywhere on the surface of a spherical boundary, the magnetic field
$B_n$ inside the normal region (given by equation \ref{toroidal-normal}) would have to be
a function of only radius at the boundary as well. This implies that we must choose a
function $h(\xi) \propto \xi^{1/2}$, so that $B_n(r,\theta) \propto \rho^{1/2}(r)$.
However, in this case, the magnetic potential becomes $\psi_n(\xi) \propto \ln\xi$
(equation \ref{toroidal-normal2}), which diverges whenever $\xi = \rho r^2\sin^2\theta$
is zero. In other words, it diverges at the center of the star ($r \to 0$), at the
surface ($\rho \to 0$), and along the symmetry axis ($\theta \to 0$). We also note that
when the magnetic induction $B_s$ inside the superconducting region (given by equation
\ref{toroidal-typeII}) is chosen so that it is angle independent (i.e. $f(\zeta) \propto
1/\zeta$), the corresponding potential is also logarithmic, $\psi_{\rm I}(\zeta) \propto
\ln\zeta$.

We therefore conclude that continuous toroidal fields, or more generally,
angle-independent magnetic inductions, are inconsistent under the assumption that $H =
H(\rho)$ holds up to the boundaries of the superconducting region. In a more realistic
treatment, $H(\rho,B)$ should be allowed to decrease smoothly to about $B_s$ near the
boundaries, which would remove the need for surface currents.

\subsection{Boundary Conditions}
Hydrostatic equilibrium for a fluid with a barotropic equation of state, in the absence
of magnetic fields, is spherically symmetric and is given by (from equation
\ref{eulerh}),
        \beq
        \m\nabla (h + \phi) = 0 \ .
        \enq
When a magnetic force that is small in comparison to pressure and gravity is applied, the
equilibrium quantities are changed by small amounts $\delta p$, $\delta\rho$, $\delta h$
and $\delta\phi$, where $\delta$ denotes Eulerian changes. Writing the magnetic force in
terms of the magnetic potential, $\m{f}_{\rm mag} = -\rho\m\nabla\psi$, the equation for
the perturbations around the background equilibrium can be written as,
        \beq
        \m\nabla(\delta h + \delta\phi + \psi) = 0 \ .
        \enq
From here it follows that,
        \beq
        \delta h = \frac{dh}{d\rho}\delta\rho = \mathfrak{B}_o - \delta\phi - \psi \ .
        \label{bernoulli}
        \enq
$\mathfrak{B}_o$ is Bernoulli's constant and is the same for the entire star. This can be
understood by treating the entire star as a single fluid region, with a magnetic
potential that varies continuously throughout the interior, but that has steep changes in
some small intervals corresponding to the boundaries.

While the background quantities $p$, $\rho$ and $\phi$ are continuous throughout the
star, their perturbations are not. Only $\delta\phi$ and its gradient are required to be
continuous, since there cannot be delta functions in mass. This implies that there will
be a density perturbation jump at a boundary, given by (from equation \ref{bernoulli}),
        \beq
        \frac{dh}{d\rho}(\delta\rho_s - \delta\rho_n) = - \psi_s + \psi_n \ .
        \label{densityjump}
        \enq
Here the subscripts $s$ and $n$ refer to the superconducting and normal regions,
respectively.

There must be substantial surface currents at the boundaries of the superconducting
shell, and therefore, the magnetic field is discontinuous across them. Otherwise, as
discussed before, the magnetic potentials become singular. From the continuity of stress,
it follows that,
        \beq
        n_j\Sigma_{ij,s} = n_j\Sigma_{ij,n} \ .
        \label{stressbc}
        \enq
$\Sigma_{ij}$ is the total stress tensor and $n_j$ is the normal unit vector of the
boundary, which in this case is simply the radial unit vector $\hr$. Thus, we require the
$rr$, $r\theta$ and $r\phi$ components of the stress tensor to be continuous. The last
two vanish identically for fluids with toroidal fields.

The total stress is,
        \beq
        \Sigma_{ij} = - \delta p \, \delta_{ij} + \sigma_{ij} \ ,
        \enq
and from equation (\ref{stressbc}), we have,
        \beq
        - \delta p_s + \sigma_{rr,s} = - \delta p_n + \sigma_{rr,n} \ .
        \label{stressbctor}
        \enq
Using the fact that for a polytrope $p = \kappa\rho^\gamma$, we have $dh/d\rho = \gamma p
/\rho^2$ and $\delta p = (\gamma p/\rho) \delta\rho$, we can combine this result with
equation (\ref{densityjump}) to get,
        \beq
        \frac{\gamma p}{\rho}(\delta\rho_s - \delta\rho_n) = - \rho(\psi_s - \psi_n)
        = \sigma_{rr,s} - \sigma_{rr,n} \ .
        \enq
The components of the stress tensor inside the normal and superconducting regions are
given by (equations \ref{stressnormal} and \ref{stresstypeII}),
        \beq
        \sigma_{rr,n} = - \frac{B_n^2}{8\pi}
        \mtext{and}
        \sigma_{rr,s} = - \rho\frac{\pa F}{\pa\rho} = - \rho\psi_{\rm II} \ .
        \label{toroidal-stress}
        \enq
Using $\psi_s = \psi_{\rm I} + \psi_{\rm II}$ (equation \ref{pottypeII}), we thus obtain,
        \beq
        - \psi_{\rm I} = - \psi_n + \frac{B_n^2}{8\pi\rho} \ .
        \label{boundary}
        \enq
This equation needs to be satisfied by the magnetic fields at the boundary. Note that
since $\psi_{\rm I} \propto H B_s/\rho$ and $\psi_n \propto B_n^2/\rho$, this equation
implies that $B_n \propto (HB_s)^{1/2}$. If we take $H \gg B_s$ to hold at the boundaries
of the superconductor as well as its interior, then the boundary condition clearly
requires $B_n \gg B_s$. Taking a more general $H(\rho,B)$, varying continuously from
$H_{\rm c1}(\rho)$ to $B_s$ through a thin boundary layer, would result in a smooth but
similar growth in the magnetic induction between the strongly type II and normal regions.
(Surface currents would be smoothed out over this boundary layer.) For entirely normal
conductors, the corresponding boundary condition simply implies the continuity of
magnetic fields.

In a more sophisticated treatment of the transitions from superconducting to normal
and/or fluid to crust, two dimensionless ratios characterize the superconducting state.
One is,
        \beq
        \kappa = \frac{\lambda}{\xi}
        \approx \frac{8.2 \Delta({\rm MeV})}{(n_{p,37})^{5/6}} \ ,
        \enq
where $\lambda$ is the London penetration depth, $\xi$ is the coherence length in the
proton superconductor, $n_p = 10^{37}n_{p,37}\,{\rm cm}^{-3}$ is the proton number
density, and $\Delta$ is the proton superconducting gap. The other is,
        \beq
        \frac{a}{\lambda} \approx 68 B_{12}^{-1/2}(n_{p,37})^{1/2} \ ,
        \enq
where $a$ is the spacing between flux tubes (Tinkham 1975). In a type II superconductor,
$\kappa>1/\sqrt{2}$.

At the crust-core boundary, $n_p$ falls dramatically, and $a/\lambda$ drops, which means
that interactions between flux tubes become important. As a result, our approximation
that $H \approx H_{\rm c1}(\rho)$ must fail, and must be replaced by a more general (and
complicated) function of both $\rho$ and $B$.

At the inner boundary of the superconducting layer, $\Delta$ ultimately disappears, and
$\kappa$ falls below $1/\sqrt{2}$. In this regime, we expect a boundary layer of a type I
superconductor to form. In fact, it is also possible for such a layer to form at the
crust-core boundary, since the gap depends exponentially on the density of states near
the proton Fermi surface, which falls with proton density. Thus, at both boundaries, we
expect the magnetic field to decrease rapidly from $H\sim 10^{15}\, \rm G$ to $B_n \sim
(HB_s)^{1/2}$.

\subsection{Derivation of the Magnetic Fields}
We will assume a simple power law relation between the magnetic field in the
superconducting region and mass density,
        \beq
        H = H_c\left(\frac{\rho}{\rho_c}\right)^\sigma \ ,
        \label{Hcrho}
        \enq
where $H_c$ and $\rho_c$ stand for the central values of the corresponding quantities.
When the superconducting region is confined to a shell, we can take $H_c$ to be the
extrapolated field strength at the center. In reality, in a strongly type II
superconductor, $H$ depends on the superconducting energy gap $\Delta$, in addition to
the proton number density $n_p$ (Tinkham 1975; Easson \& Pethick 1977). Both $n_p$ and
$\Delta$ are functions of baryon density $\rho$ (Elgar{\o}y et al. 1996; Baldo \& Schulze
2007). $\Delta$ vanishes at sufficiently high densities, and protons become normal. At
low densities, superconductivity is suppressed since protons are bound in the nuclei in
the neutron star crust. In both cases, the transition from superconducting to normal
state may be sharp and we take the form given by equation (\ref{Hcrho}) in
superconducting regions.

In this case, equations (\ref{psiI}) and (\ref{psiII}) imply $\psi_{\rm I} = g(\zeta)$
and $\psi_{\rm II} = \sigma\zeta g'(\zeta)$, where $\zeta = Hr\sin\theta$. Consider a
power law function of the form $g(\zeta) = N\zeta^n$, where $N$ is a constant; then
$\psi_{\rm I} = N\zeta^n$ and $\psi_{\rm II} = n\sigma N\zeta^n$, so that the total
magnetic potential becomes,
        \beq
        \psi_s = \psi_{\rm I} + \psi_{\rm II} = (n\sigma+1)N\zeta^n \ .
        \enq
We exclude $n = 0$ since that corresponds to zero magnetic induction and force. On the
other hand, for $n < 0$ the magnetic potential diverges when either $r\to 0$ or
$\theta\to 0$. Moreover, the magnetic force diverges in the same limits in the interval
$0 < n < 1$ . Therefore, the only nonsingular choices are $n \geqslant 1$. The magnetic
induction inside the superconductor is (equation \ref{toroidal-typeII}),
        \beq
        B_s(r,\theta) = B_o
        \left(\frac{\rho}{\rho_c}\right)^{\sigma(n-1)+1}
        \left(\frac{r}{r_o}\right)^n\sin^n\theta
        \mtext{where}
        B_o = 4\pi n N\rho_c {H_c}\!^{n-1}{r_o}\!^n \ .
        \label{toroidal-typeIIB}
        \enq
The constant $r_o$ will be defined later. The corresponding magnetic potential can be
written as,
        \beq
        \psi_s(r,\theta) = \Psi_o \left(\frac{\rho}{\rho_c}\right)^{n\sigma}
        \left(\frac{r}{r_o}\right)^n \sin^n\theta \mtext{where}
        \Psi_o = \frac{(n\sigma+1)H_cB_o}{4\pi n\rho_c} \ .
        \label{toroidal-typeIIpot}
        \enq

Inside the normal region we have, from equations (\ref{toroidal-normal}) and
(\ref{toroidal-normal2}), defining $\xi = \rho r^2\sin^2\theta$,
        \beq
        B_n(r,\theta) = \frac{h(\xi)}{r\sin\theta} \mtext{and}
        \psi_n'(\xi) = \frac{h(\xi)h'(\xi)}{4\pi\xi} \ .
        \enq
We will assume a power law for the arbitrary function, $h(\xi) = M\xi^m$, where $M$ is a
constant. Then,
        \beq
        \frac{B_n^2}{8\pi\rho} = \frac{M^2\xi^{2m - 1}}{8\pi}
        \mtext{and}
        \psi_n = \frac{mM^2\xi^{2m-1}}{4\pi(2m-1)} \ .
        \enq
The boundary condition (equation \ref{boundary}) gives, after some rearrangement,
        \beq
        N\zeta^n = \frac{M^2\xi^{2m-1}}{8\pi(2m-1)} \ .
        \enq
In order to satisfy this equation for all values of $\theta$ at the boundary (which we
will assume to be located at some radius $r = r_b$) we must have,
        \beq
        n = 4m - 2 \mtext{whence}
        M = \left[\frac{4\pi n NH^n(r_b)}{\rho^{n/2}(r_b)}\right]^{1/2} \ .
        \enq
Then the magnetic field in the normal region is,
        \beq
        B_n(r,\theta) = \hat B_o \left(\frac{\rho}{\rho_c}\right)^{(n+2)/4}
        \left(\frac{r}{r_o}\right)^{n/2}\sin^{n/2}\theta \mtext{where}
        \hat B_o = M\rho_c^{(n+2)/4}r_o^{n/2} \ .
        \label{toroidal-normalB}
        \enq
Note that $B_s$ and $B_n$ must have different angular dependencies in order for the
potentials $\psi_s$ and $\psi_n$ to be consistent. Moreover,
        \beq
        \hat B_o
        = (H_cB_o)^{1/2}\left[\frac{\rho(r_b)}{\rho_c}\right]^{n(2\sigma - 1)/4} \ ,
        \label{hatBo}
        \enq
so that the magnetic fields in the normal regions are moderately strong. The magnetic
potential in the normal region is,
        \beq
        \psi_n(r,\theta) = \hat \Psi_o \left(\frac{\rho}{\rho_c}\right)^{n/2}
        \left(\frac{r}{r_o}\right)^n\sin^n\theta \mtext{where}
        \hat \Psi_o
        = \frac{(n+2)\hat B_o^2}{8\pi n\rho_c} \ .
        \label{toroidal-normalpot}
        \enq
Thus, it follows that $\hat\Psi_o \propto \Psi_o$,
        \beq
        \frac{\hat\Psi_o}{\Psi_o} =
        \frac{n+2}{2(n\sigma+1)}\frac{\hat B_o^2}{H_cB_o} =
        \frac{n+2}{2(n\sigma+1)}\left[\frac{\rho(r_b)}{\rho_c}\right]^{n(2\sigma - 1)/2} \ .
        \label{toroidal-ratio}
        \enq
As in the superconducting case, we need to have $n \geqslant 1$ in order to avoid any
divergences in the potentials or forces.

\subsection{The $n = 1$ Case}
In a later section, we will show that toroidal fields by themselves are unstable, and
that the $n = 1$ case is the closest to being stable. We will be concerned particularly
with cases where $H\propto\rho$, i.e. $\sigma = 1$. This corresponds to taking the proton
number density to be proportional to the baryon density, $n_p\propto\rho$, and neglecting
logarithmic dependencies in $H$, which is a good first order approximation (Easson \&
Pethick 1977; Muzikar \& Pethick 1981). The magnetic potentials in the superconducting
and normal regions become, from equations (\ref{toroidal-typeIIpot}) and
(\ref{toroidal-normalpot}),
        \beq
        \psi_s = \Psi_o\left(\frac{\rho}{\rho_c}\right)
        \left(\frac{r}{r_o}\right)\sin\theta
        \mtext{and}
        \psi_n = \hat\Psi_o\left(\frac{\rho}{\rho_c}\right)^{1/2}
        \left(\frac{r}{r_o}\right)\sin\theta \ ,
        \label{n1case}
        \enq
where, from equation (\ref{toroidal-ratio}), we have,
        \beq
        \Psi_o = \frac{H_cB_o}{2\pi\rho_c} \mtext{and}
        \frac{\hat\Psi_o}{\Psi_o} =
        \frac{3}{4}\left[\frac{\rho(r_b)}{\rho_c}\right]^{1/2} \ .
        \label{n1case2}
        \enq
The angular part of the potentials can be expanded in Legendre polynomials,
        \beq
        \sin\theta = \sum_{\ell = 0}^{\infty} \Theta_\ell P_\ell(\cos\theta) \ .
        \enq
Only even $\ell$ remain in the series and the coefficients are,
        \beq
        \Theta_\ell = \frac{2\ell + 1}{2}\int_{-1}^{1}
        \sin\theta P_\ell(\cos\theta)d(\cos\theta)
        = \frac{(2\ell + 1)\pi^2}{2(\ell + 2)(1 - \ell)\Gamma^2(\ell/2 + 1)
        \Gamma^2(1/2 - \ell/2)} \ .
        \enq
In particular, $\Theta_0 = \pi/4$. Subsequent terms in the expansion have the ratio,
        \beq
        \frac{\Theta_{\ell + 2}}{\Theta_\ell} =
        \frac{(2\ell + 5)(\ell + 1)(\ell - 1)}
        {(2\ell + 1)(\ell + 4)(\ell + 2)} \ .
        \enq
Clearly, $\Theta_{\ell + 2}/\Theta_\ell\to 1$ as $\ell\to\infty$. The result can also be
expressed in terms of the spherical harmonics which are related to the Legendre
polynomials through,
        \beq
        Y_\ell(\theta) = \sqrt{\frac{2\ell + 1}{4\pi}} P_\ell(\cos\theta) \ .
        \enq
Then, for even $\ell$,
        \beq
        \sin\theta = \sum_{\ell = 0}^{\infty} \tilde\Theta_\ell Y_\ell(\theta)
        \mtext{where} \tilde\Theta_\ell = \sqrt{\frac{4\pi}{2\ell + 1}}\Theta_\ell \ .
        \enq

        \begin{figure}
        \centerline{\includegraphics[scale=1.1]{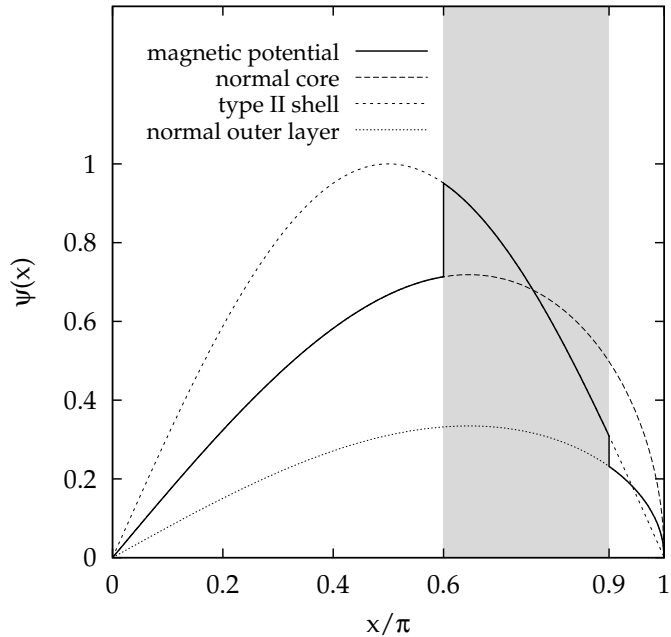}}
        \caption[Magnetic potential profile for a three component star.]
        {Magnetic potential profile for a three component star with a normal core,
        type II superconducting shell, and surrounding normal layer. The potential is shown for
        the $n = \sigma = 1$ case for the magnetic field (equation \ref{n1case}), and a $\gamma = 2$
        polytropic equation of state. The superconducting shell lies between $x_1 = 0.6\pi$ and
        $x_2 = 0.9\pi$, and is shown shaded. The potential is shown along the equator of the star,
        i.e. $\sin\theta = 1$, in units of $\Psi_o$ defined in equation (\ref{n1case2}). The
        profiles for the potentials within each region are shown extended over the whole star
        for comparison.}
        \label{figurepsi}
        \end{figure}

We will consider a $\gamma = 2$ polytrope for which the equation of state is $p =
\kappa\rho^2$, where $\kappa$ is a constant. In this case, the background density is of
the form $\rho = \rho_c\sin x/x$, in terms of the dimensionless variable $x = r/r_o$,
where $r_o = \sqrt{\kappa/2\pi G}$. The stellar radius is $R_\star = \pi r_o$, and the
stellar mass is $M_\star = \pi M_o$, where $M_o = 4\pi\rho_c r_o^3$. The central density
is given by $\rho_c = \pi M_\star/4R_\star^3$. For a neutron star with $M_\star \approx
1.4M_\odot$ and $R_\star \approx 10^6\,\rm cm$, we have $\rho_c \approx 2.2\times
10^{15}\,\rm g/cm^3$.

As noted before, superconductivity exists only within a certain range of densities, or
equivalently, a range of radii, which we will denote by $x_1 < x < x_2$. In particular,
it is suppressed in the crust where the protons become bound in nuclei. The crust exists
at densities below $\rho \approx 2\times 10^{14}\,\rm g/cm^3$ (Baym et al. 1971; Lorenz
et al. 1993), corresponding to an outer radius of $x_2 \approx 0.9 \pi$. On the other
hand, the proton pairing gap vanishes at higher densities. This cutoff for
superconductivity is not as well-established and estimates range from $\rho \approx
5\times 10^{14}\,\rm g/cm^3$ to $10^{15}\,\rm g/cm^3$ (Elgar{\o}y et al. 1996; Baldo \&
Schulze 2007). Thus, the inner boundary of the superconducting shell ranges from $x_1
\approx 0.8\pi$ to $0.6\pi$, respectively.

The magnetic potential for the $n = 1$ case in a three component star consisting of a
type II superconducting shell surrounded by normal regions (as depicted in
fig.~\ref{sketch}) is shown in fig.~\ref{figurepsi}. Note that the potential within the
superconducting shell (which is taken to be in the interval $0.6 < x/\pi < 0.9$) is
larger than those in the normal regions.

\subsection{Calculation of the Gravitational Potential Perturbation}
The gravitational potential perturbations are given by the perturbed Poisson's equation,
        \beq
        \nabla^2\delta\phi = 4\pi G\delta\rho \ .
        \label{poissontypeII}
        \enq
For a $\gamma = 2$ polytrope, we have $dh/d\rho = p'(\rho)/\rho = 2\kappa$, and equation
(\ref{bernoulli}) becomes $2\kappa\delta\rho = \mathfrak{B}_o - \delta\phi - \psi$.
Expanding the perturbations in spherical harmonics as $\delta\phi(x,\theta) =
\phi_\ell(x) Y_\ell(\theta)$ and so on, Poisson's equation gives,
        \beq
        \frac{1}{x^2}\frac{d}{dx}\left(x^2\frac{d\phi_\ell}{dx}\right)
        + \left[1 - \frac{\ell(\ell + 1)}{x^2}\right]\phi_\ell
        = \mathfrak{B}_o\delta_{\ell 0} - \psi_\ell \ .
        \enq

The complete solution of this equation is the sum of a homogeneous solution and a
particular solution. The homogeneous solution is given in terms of the spherical Bessel
functions, $\phi_h(x) = A_\ell j_\ell(x) + B_\ell y_\ell(x)$, and the particular solution
can be found by the method of variation of parameters, $\phi_p(x) = \tilde A_\ell(x)
j_\ell(x) + \tilde B_\ell(x) y_\ell(x)$. Thus, the gravitational potential perturbations
in the three regions (core, superconducting shell and outer normal layer, as depicted in
fig.~\ref{sketch}) are,
        \beq
        \begin{array}{l}\vspace{0.2cm}\displaystyle
        \phi_{c,\ell}(x) =
        \left[A_\ell + \tilde A_\ell(x)\right] j_\ell(x) +
        \left[B_\ell + \tilde B_\ell(x)\right] y_\ell(x) + \mathfrak{B}_o\delta_{\ell 0}
        \\ \vspace{0.2cm}\displaystyle
        \phi_{s,\ell}(x) =
        \left[C_\ell + \tilde C_\ell(x)\right] j_\ell(x) +
        \left[D_\ell + \tilde D_\ell(x)\right] y_\ell(x) + \mathfrak{B}_o\delta_{\ell 0}
        \\ \displaystyle
        \phi_{n,\ell}(x) =
        \left[E_\ell + \tilde E_\ell(x)\right] j_\ell(x) +
        \left[F_\ell + \tilde F_\ell(x)\right] y_\ell(x) + \mathfrak{B}_o\delta_{\ell 0}
        \end{array}
        \enq
where $A_\ell$ through $F_\ell$ are constants, and we define,
        \beq
        \tilde A_\ell(x) = - \int_{x}^{\pi} t^2 \psi_{c,\ell}(t) y_\ell(t) dt \mtext{and}
        \tilde B_\ell(x) = - \int_{0}^{x} t^2 \psi_{c,\ell}(t) j_\ell(t) dt \ .
        \enq
Here $\psi_{c,\ell}$ refers to the $\ell$-th component of the spherical harmonic
expansion of the potential $\psi_c$. The remaining coefficients are defined in an
analogous fashion. Note that the integration boundaries can be arbitrarily adjusted,
which amounts to a redefinition of the constants $A_\ell$ through $F_\ell$ above. The
particular choice made here makes sure there are no singularities, but is otherwise
immaterial.

Since there can be no gravitational forces in the center, the gradient of the
gravitational potential must vanish there. This implies that as $x\to 0$ we must have
$\phi_\ell\to$ constant for $\ell = 0$, and $\phi_\ell\to 0$ and $\phi_\ell'\to 0$ for
$\ell\ne 0$. As $x \to 0$, the limiting values of the spherical Bessel functions are
$j_\ell \propto x^\ell$ and $y_\ell \propto x^{- \ell - 1}$. It therefore follows that
$B_\ell = 0$ for all values of $\ell$. The remaining five coefficients $A_\ell$,
$C_\ell$, $D_\ell$, $E_\ell$ and $F_\ell$, and Bernoulli's constant $\mathfrak{B}_o$ are
to be determined from the continuity of the potentials and their derivatives across the
shell boundaries, which we will take to be located at $x_1$ and $x_2$, such that $x_1 <
x_2$,
        \beq
        \begin{array}{c}\vspace{0.2cm}\displaystyle
        \phi_{c,\ell}(x_1) = \phi_{s,\ell}(x_1) \mtext{and}
        \phi_{c,\ell}'(x_1) = \phi_{s,\ell}'(x_1)
        \\ \displaystyle
        \phi_{s,\ell}(x_2) = \phi_{n,\ell}(x_2) \mtext{and}
        \phi_{s,\ell}'(x_2) = \phi_{n,\ell}'(x_2)
        \end{array}
        \label{toroidal-bc1}
        \enq
and from the boundary conditions at the stellar surface, which is located at $x = \pi$,
        \beq
        \begin{array}{ll}\vspace{0.2cm}\displaystyle
        \pi\phi_{n,\ell}'(\pi) + (\ell + 1)\phi_{n,\ell}(\pi) = 0 & \mtext{for} \ell \ne 0
        \\ \displaystyle \phi_{n,\ell}'(\pi) = \phi_{n,\ell}(\pi) = 0 & \mtext{for} \ell = 0
        \end{array}
        \label{toroidal-bc2}
        \enq
The surface boundary conditions follow from the multipole expansion of the gravitational
potential, which implies that $\phi_\ell \propto x^{- \ell -1}$, and the conservation of
mass, which additionally implies $\phi_\ell = 0$ for $\ell = 0$.

        \begin{figure}
        \centerline{\includegraphics[scale=1]{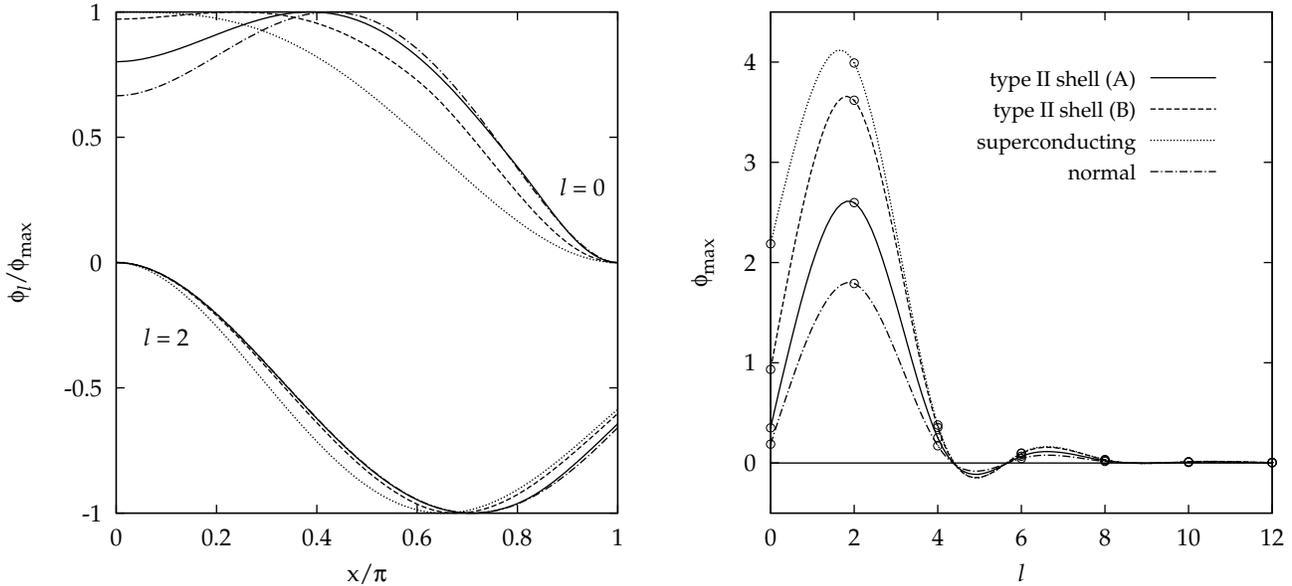}}
        \caption[Gravitational potential perturbation for a fluid star with toroidal fields.]
        {Gravitational potential perturbation for a fluid star with toroidal
        fields, expanded in spherical harmonics for the $n=1$ case (equation \ref{n1case}). The
        potentials are shown for four sample models: type II superconducting shell
        between $x_1 = 0.8\pi$ and $x_2 = 0.9\pi$ (case A) and between $x_1 = 0.6\pi$ and
        $x_2 = 0.9\pi$ (case B), completely superconducting star ($x_1 = 0$ and $x_2 = \pi$),
        and completely normal star ($x_1 = x_2 = 0.9\pi$).
        The figure on the left shows the first two harmonics $\phi_\ell$ (for $\ell = 0$ and $\ell = 2$)
        scaled by the maximum value of the potential, $\phi_{\rm max}$. The figure on the
        right shows $\phi_{\rm max}$ for the first few $\ell$, in units of $\Psi_o$ defined in
        equation (\ref{n1case2}). The points for different values of $\ell$ (shown with circles)
        are connected by a cubic spline curve. The amplitude of $\phi_\ell$ decreases
        sharply with $\ell$.}
        \label{figurephi}
        \end{figure}

        \begin{figure}
        \centerline{\includegraphics[scale=1]{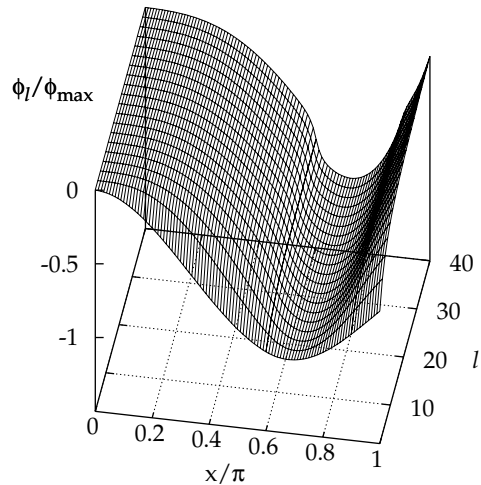}}
        \caption[Gravitational potential perturbation for a fluid star as a function of $\ell$.]
        {Gravitational potential perturbation for a fluid star as a function of
        $\ell$. The potential is shown for the $n = 1$ case of a three component star with a
        superconducting shell between $x_1 = 0.6\pi$ and $x_2 = 0.9\pi$. The same scaling
        is used as in fig.~\ref{figurephi}, and only $\ell > 0$ are shown.}
        \label{figurephi3D}
        \end{figure}

Making use of various relations between spherical Bessel functions,\footnote{In
particular, letting $f_\ell$ denote either $j_\ell$ or $y_\ell$, we have
$j_\ell(x)y_\ell\!'(x) - j_\ell\!'(x)y_\ell(x) = x^{-2}$,
$xf_\ell\!'(x) = xf_{\ell - 1}(x) - (\ell + 1)f_\ell(x)$ and
$(2\ell+1)f_\ell\!'(x) = \ell f_{\ell-1}(x) - (\ell+1)f_{\ell+1}(x)$.}
the continuity conditions at the shell boundaries (equation \ref{toroidal-bc1}) yield,
        \beq
        \begin{array}{l}\vspace{0.2cm}\displaystyle
        A_\ell + \tilde A_\ell(x_1) = C_\ell + \tilde C_\ell(x_1)
        \mtext{and} \tilde B_\ell(x_1) = D_\ell + \tilde D_\ell(x_1)
        \\ \displaystyle
        C_\ell + \tilde C_\ell(x_2) = E_\ell + \tilde E_\ell(x_2)
        \mtext{and} D_\ell + \tilde D_\ell(x_2) = F_\ell + \tilde F_\ell(x_2)
        \end{array}
        \enq
and the surface boundary conditions (equation \ref{toroidal-bc2}) give, since $\tilde
E_\ell(\pi) = 0$,
        \beq
        \begin{array}{ll}\vspace{0.2cm}\displaystyle
        E_\ell j_{\ell-1}(\pi) + \left[F_\ell + \tilde F_\ell(\pi)\right] y_{\ell - 1}(\pi) = 0
        & \mtext{for} \ell \ne 0 \\ \displaystyle
        \mathfrak{B}_o = \frac{E_\ell}{\pi^2y_1(\pi)} = - \frac{F_\ell + \tilde F_\ell(\pi)}{\pi^2j_1(\pi)}
        & \mtext{for} \ell = 0 \end{array}
        \enq

Special cases can be considered. For instance, for $x_1 = 0$ and $x_2 = \pi$ we retrieve
the completely superconducting star. In this case $\tilde B_\ell(x_1) = \tilde
D_\ell(x_1) = 0$ so that $D_\ell = 0$. Since $\tilde C_\ell(x_2) = 0$ as well, the
surface boundary conditions reduce to,
        \beq
        \begin{array}{ll}\vspace{0.2cm}\displaystyle
        C_\ell j_{\ell - 1}(\pi) + \tilde D_\ell(\pi)y_{\ell - 1}(\pi) = 0
        & \mtext{for} \ell \ne 0
        \\ \displaystyle
        \mathfrak{B}_o = \frac{C_\ell}{\pi^2 y_1(\pi)} = - \frac{\tilde D_\ell(\pi)}{\pi^2 j_1(\pi)}
        & \mtext{for} \ell = 0
        \end{array}
        \enq
On the other hand, letting $x_1 \to 0$ while keeping $x_2 < \pi$ we retrieve the case of
a superconducting core surrounded by a normal region. When $x_1 = x_2$ the star is
completely normal conducting. All such cases are equivalent, up to a scaling determined
by the magnitude of the magnetic potential (which is given through equation
\ref{toroidal-ratio}). Sample models are shown in figs.~\ref{figurephi} and
\ref{figurephi3D} for the $n = 1$ case discussed before (equation \ref{n1case}).

\subsection{Density Perturbation}
\begin{figure}
        \centerline{\includegraphics[scale=1]{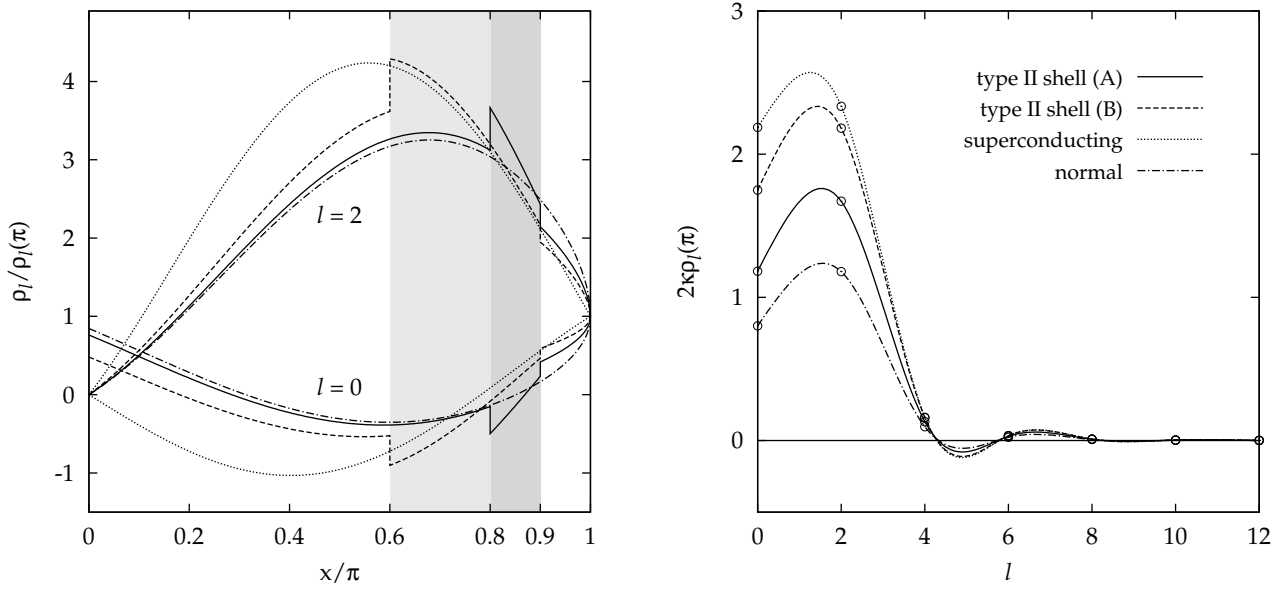}}
        \caption[Density perturbations for a fluid star with toroidal fields.]
        {Density perturbations for a fluid star with toroidal fields, expanded
        in spherical harmonics. Plots are shown for the same four sample cases considered
        in fig.~\ref{figurephi}. The figure on the left shows the first two harmonics
        $\rho_\ell$ (for $\ell = 0$ and $\ell = 2$) scaled by the surface value of the
        density perturbation $\rho_\ell(\pi)$. The shaded regions indicate the position
        of the superconducting shell. The figure on the right shows $2\kappa\rho_\ell(\pi)$
        for the first few $\ell$,  in units of $\Psi_o$ defined in equation (\ref{n1case2}).}
        \label{figurerho}
        \end{figure}

The density perturbation within each region can be calculated through equation
(\ref{bernoulli}), which for a $\gamma = 2$ polytrope becomes,
        \beq
        2\kappa\delta\rho = \mathfrak{B}_o - \delta\phi - \psi \ .
        \enq
Sample plots of density perturbations for the $n=1$ case are shown in
fig.~\ref{figurerho}. The density jump at a boundary is then given through,
        \beq
        2\kappa\Delta_\rho = 2\kappa(\delta\rho_{\rm in} - \delta\rho_{\rm out})
        = \psi_{\rm out} - \psi_{\rm in} \ .
        \label{toroidaljumpeq}
        \enq
In particular, consider the density jump when going from a normal region into a
superconducting region at a boundary $r = r_b$. Using equations
(\ref{toroidal-typeIIpot}) and (\ref{toroidal-normalpot}), we get,
        \beq
        2\kappa\Delta_\rho = 2\kappa(\delta\rho_n - \delta\rho_s) = \psi_s - \psi_n
        = \frac{n(2\sigma - 1)}{2(n\sigma+1)}\psi_s(r_b,\theta)
        \mtext{where} n \geqslant 1 \ .
        \enq
Note that $\Delta_\rho \geqslant 0$ for $\sigma > 1/2$. In other words, the density
perturbation \emph{decreases} when going from a normal region into a superconducting
region, and vice versa. Also note that the jump goes to zero at the poles, i.e.
$\Delta_\rho\to 0$ as $\theta\to 0$, since the magnetic potentials vanish there.

The relation between the Eulerian density perturbation and the Lagrangian displacement is
given through,
        \beq
        \delta\rho = - \m\nabla\cdot(\rho\m\xi) = - \rho\m\nabla\cdot\m\xi - \rho'\xi_r \ .
        \enq
Normally, the term $\m\nabla\cdot\m\xi$ inside the fluid is undetermined. However, at the
surface $\rho = 0$, so that we can calculate the radial displacement, which determines
the shape of the perturbed stellar surface,
        \beq
        \xi_r = - \delta\rho/\rho' \ .
        \enq
For a $\gamma = 2$ polytrope we have $\rho = \rho_c\sin x/x$, so that at the surface
$\rho'(\pi) = - \rho_c/\pi$ and $\xi_r = \pi\delta\rho/\rho_c$. The $\ell = 0$ term in
the spherical harmonic expansion of $\xi_r$ defines a spherically symmetric expansion (or
compression) of the star, while higher order $\ell$ determine the deformation of the
surface as a function of the polar angle, $\theta$.

\subsection{Quadrupolar Distortion}
The moment of inertia of the unperturbed star is given by,
        \beq
        I_{ij} = \int_V \rho(r^2 \delta_{ij} - r_ir_j) \, d^3 r \ .
        \enq
Since the star is initially spherically symmetric we have $I_{xx} = I_{yy} = I_{zz}$. For
a $\gamma = 2$ polytrope the density profile is given through $\rho = \rho_c\sin x/x$, so
that the moment of inertia becomes,
        \beq
        I_o \equiv I_{xx} = \int_V \rho r^2\left(1 - \sin^2\theta\cos^2\varphi\right) d^3r
        = \frac{8(\pi^2 - 6)\rho_cR_\star^5}{3\pi^3} \ .
        \enq
Here $R_\star$ is the stellar radius, which corresponds to $x = R_\star/r_o = \pi$.

\begin{figure}
        \centerline{\includegraphics[scale=1]{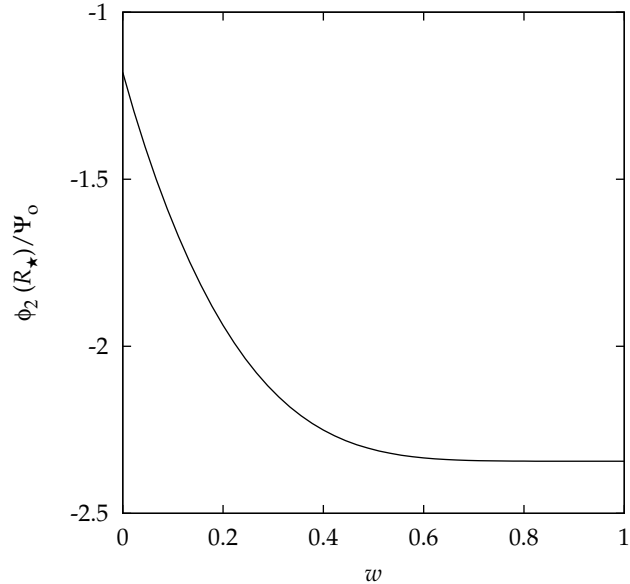}}
        \caption[$\phi_2(R_\star)$ as a function of the width of the superconducting shell.]
        {$\phi_2(R_\star)$ as a function of the width $w = (x_2 - x_1)/x_2$ of the
        superconducting shell in a three component star. The outer radius of the shell is
        fixed at $x_2 = 0.9\pi$. The type II shell models (cases A and B) listed in table
        \ref{tablephi} are retrieved by setting $x_1 = 0.8\pi$ ($w = 1/9$) and $x_1 =
        0.6\pi$ ($w = 1/3$), respectively. When $x_1 = x_2$ ($w = 0$) the star becomes
        normal.}
        \label{figurequadrupole}
        \end{figure}

The application of the magnetic perturbation renders the star axisymmetric ($I_1 = I_2
\ne I_3$). In this case the moments of inertia become $I_1 = I_o + \delta I_1$ around an
axis that lies in the equatorial plane, and $I_3 = I_o + \delta I_3$ around the axis of
symmetry which passes through the poles. We will define the star to be \emph{oblate} when
$\delta I_3 > \delta I_1$ and \emph{prolate} when $\delta I_3 < \delta I_1$. In other
words, when more of the mass is distributed towards the equator the star is oblate, and
when more of the mass is closer to the poles the star is prolate. The difference between
the moments of inertia is related to the \emph{gravitational quadrupole moment}, which in
turn is related to the $\ell = 2$ harmonic of the gravitational potential at the stellar
surface,
        \beq
        Q_{20} = \int_V \rho r^2 Y_2(\theta)d^3 r
        = - \sqrt{\frac{5}{4\pi}}(\delta I_3 - \delta I_1)
        = - \frac{5R_\star^3\phi_2(R_\star)}{4\pi G} \ .
        \enq
Thus,
        \beq
        \phi_2(R_\star) =
        \sqrt{\frac{4\pi}{5}}\frac{G(\delta I_3 - \delta I_1)}{R_\star^3} \ .
        \enq
Therefore, the sign of $\phi_2$ at the surface determines whether the star is prolate or
oblate. Note that for all the cases shown in fig.~\ref{figurephi}, $\phi_2(R_\star)$ is
negative and consequently the star is prolate. The precession frequency of an
axisymmetric star is $\sim\epsilon\Omega_\star$, where $\Omega_\star$ is the angular
velocity and $\epsilon$ is a dimensionless constant defined through,
        \beq
        \epsilon = \frac{I_3 - I_1}{I_1} \approx \frac{\delta I_3 - \delta I_1}{I_o}
        = \frac{3\pi^2\sqrt{5\pi}\phi_2(R_\star)}{16(\pi^2 - 6) G\rho_c R_\star^2} \ .
        \enq
For the $n=1$ case, the gravitational potential perturbations are measured in units of
$\Psi_o = H_cB_o/2\pi\rho_c$ (equation \ref{n1case2}). The central density for a $\gamma
= 2$ polytrope is $\rho_c = \pi M_\star/4R_\star^3$. Thus, we can rewrite the above
equation as,
        \beq
        \epsilon = 0.945\times 10^{-9} \left(\frac{\phi_2(R_\star)}{\Psi_o}\right)
        \left(\frac{H_c}{10^{15} \, \rm G}\right)
        \left(\frac{B_o}{10^{12} \, \rm G}\right)
        \left(\frac{R_\star}{10 \, \rm km}\right)^4
        \left(\frac{M_\star}{1.4M_\odot}\right)^{-2} \ .
        \label{epsilon}
        \enq
Sample values of $\phi_2(R_\star)$ are listed in table \ref{tablephi}, and
$\phi_2(R_\star)$ as a function of superconducting shell width in a three component star
is plotted in fig.~\ref{figurequadrupole}. Note that the values of $\epsilon$ for the
various models are very similar. This should not be surprising, as the magnetic fields in
all cases are of similar magnitude.

        \begin{table}
        \caption[Values of $\phi_2(R_\star)$ for the cases considered in fig.~\ref{figurephi}.]
        {Values of $\phi_2(R_\star)$ for the cases considered in fig.~\ref{figurephi}.
        The negative signs signify the fact that the models considered here are prolate,
        i.e. $\delta I_1 > \delta I_3$.}
        \label{tablephi}
        \center{\begin{tabular}{|c|c|}
        \hline Case & $\phi_2(R_\star)/\Psi_o$ \\
        \hline
        type II shell (A) & $-1.67$ \\
        type II shell (B) & $-2.18$ \\
        superconducting & $-2.33$ \\
        normal & $-1.18$ \\
        \hline
        \end{tabular}}
        \end{table}

In particular, the normal case considered here (in figs.~\ref{figurephi} and
\ref{figurerho}, and in table \ref{tablephi}) is for a magnetic field of strength $\hat
B_o = (H_cB_o)^{1/2}\left[\rho(x_2)/\rho_c\right]^{1/4} \approx 1.8\times 10^{13}\, \rm
G$ (equation \ref{hatBo}). This is simply the limiting value of the normal field as the
superconducting shell vanishes, $x_1 \to x_2$. In the normal case, the magnetic potential
is given in units of $\hat\Psi_o = 3\hat B_o^2/8\pi\rho_c$ (equation
\ref{toroidal-normalpot}), which can be evaluated for different choices of $\hat B_o$.

\section{Stability of Magnetic Fields}\label{sectionstability}
In this section, we will discuss the stability of toroidal fields in neutron stars. We
will follow the \emph{energy principle} considerations outlined in Bernstein et al.
(1958) and Tayler (1973). An extensive review is also given in Freidberg (1982). The
formalism that is developed in this section is valid for any $H(\rho,B)$ and is
applicable to both normal and superconducting neutron stars. For the purpose of this
section, we will treat the entire star as either normal or superconducting, and therefore
will not worry about internal boundaries.

We also ignore rotation, and thus do not need to pay attention to ``trivial''
displacements discussed by Friedman \& Schutz (1978). In magnetic stars, trivial modes
are defined by the requirements that $\delta\rho = 0$ and $\delta\m{B} = 0$. Since we
will express the energy of the perturbations in terms of $\delta\rho$ and $\delta\m{B}$,
trivial displacements will have no effect on it (see equation B60 in Friedman \& Schutz
1978 and footnote 3 in Glampedakis \& Andersson 2007). However, in a rotating star,
trivial displacements will have to be taken into consideration.

Glampedakis \& Andersson (2007) emphasize the importance of the magnetic field for
rotating stars by showing that sufficiently strong fields can stabilize inertial modes
that would otherwise be unstable. The same will be true for type II superconducting
stars. We will not treat rotation-induced instabilities here. Instead, we emphasize the
effects of the magnetic free energy $F(\rho,B)$ in a type II superconductor. Energy
conditions presume zero dissipation. Moreover, we consider a single fluid, which in
reality consists of at least three fluids: neutrons, protons and electrons. There will be
additional buoyant modes which may or may not alter the stability conditions we derive.

Assuming small oscillatory perturbations about equilibrium, we have, from equation
(\ref{eulertypeII}),
        \beq
        - \rho\frac{d^2\m\xi}{dt^2} = \rho\omega^2\m\xi
        = \delta \left(\m\nabla p + \rho\m\nabla\phi - \m{f}_{\rm mag}\right)
        = - \m{\cal F}(\m\xi) \ .
        \enq
The force operator $\m{\cal F}$ is self-adjoint, which implies that the eigenvalues
$\omega^2$ are real. One condition for stability is that all frequencies $\omega$ be
real, so that there are no growing modes. Alternatively, the variation in the total
potential energy due to the perturbations should always be positive,
        \beq
        \delta W = - \frac{1}{2}\int \m\xi\cdot\m{\cal F}(\m\xi)\,d V > 0 \ .
        \enq
To lowest order, the integration is carried over the equilibrium volume. The Lagrangian
and Eulerian pressure perturbations are given by $\Delta p = (\gamma p/\rho)\Delta\rho =
- \gamma p \m\nabla\cdot\m\xi$ and $\delta p = \Delta p - \m\xi\cdot\m\nabla p = - \gamma
p \m\nabla\cdot\m\xi - \m\xi\cdot\m\nabla p$. Here $\gamma$ is for the perturbations, and
in general may differ from the background polytropic index. The difference gives rise to
buoyancy terms, which will not be considered in this paper, however we will comment on
their effects on stability briefly.

Integrating by parts, we get,
        \beq
        \begin{array}{rcl}\vspace{0.2cm}
        \delta W &=& \delta W_{\rm p} + \delta W_{\rm mag} \\ \vspace{0.2cm}
        \delta W_{\rm p} &=& \displaystyle
        \frac{1}{2} \int \bra{[} \gamma p (\m\nabla\cdot\m\xi)^2 +
        (\m\xi\cdot\m\nabla p) (\m\nabla\cdot\m\xi) - (\m\xi\cdot\m\nabla\phi)
        (\m\nabla\cdot \rho\m\xi) + \rho\m\xi\cdot\m\nabla\delta\phi \ket{]} d V
        \\ && \vspace{0.2cm}\displaystyle
        - \frac{1}{2} \oint d\m{S}\cdot\m\xi \bra{[}\gamma p \m\nabla\cdot\m\xi +
        \m\xi\cdot\m\nabla p \ket{]} \\
        \delta W_{\rm mag} &=& \displaystyle
        - \frac{1}{2} \int \m\xi\cdot\delta\m{f}_{\rm mag} \, d V
        \end{array}
        \label{energy}
        \enq
We will refer to the two parts in the energy as the hydrostatic part $\delta W_{\rm p}$,
which includes the contributions from pressure and gravity, and the magnetic part $\delta
W_{\rm mag}$. In equilibrium, the pressure and density are related through a polytropic
equation of state and consequently they both go to zero at the surface. Therefore, the
surface integral vanishes.

We now turn our attention to the calculation of the magnetic energy variation. Faraday's
law gives the variation in the magnetic field in a perfect conductor as,
        \beq
        \delta\m{B} = \m\nabla\times(\m\xi\times\m{B}) \ .
        \label{magper}
        \enq
We next discuss the normal and superconducting cases separately.

\subsection{Normal Conducting Star}
In a normal conducting medium, the force is given as,
        \beq
        \m{f}_{\rm mag} = \frac{\m{J}\times\m{B}}{c}
        = \frac{(\m\nabla\times\m{B})\times\m{B}}{4\pi} \ .
        \enq
The perturbed force becomes,
        \beq
        \delta \m{f}_{\rm mag} = \frac{\delta\m{J}\times\m{B}}{c} +
        \frac{\m{J}\times\delta\m{B}}{c} \mtext{where} \frac{\delta\m{J}}{c} =
        \frac{\m\nabla\times\delta\m{B}}{4\pi} \ .
        \enq
Integrating the first term in $\delta W_{\rm mag}$, given through equation
(\ref{energy}), by parts and rearranging, we thus have,
        \beq
        \delta W_{\rm mag} = - \frac{1}{2}\int \m\xi\cdot\delta\m{f}_{\rm mag}\,d V
        = \frac{1}{2}\int \left[\frac{|\delta\m{B}|^2}{4\pi}
        - \frac{\m{J}\cdot\delta\m{B}\times\m\xi}{c}\right] d V
        + \frac{1}{8\pi}\oint d\m{S}\cdot\bra{[}\m\xi(\m{B}\cdot\delta\m{B})
        - \m{B}(\m\xi\cdot\delta\m{B})\ket{]} \ .
        \label{stabilitynormal}
        \enq
The first surface integral vanishes when $d\m{S}\cdot\m{B} = 0$, i.e. when the magnetic
field is perpendicular to the surface, as is the case for a toroidal field. On the other
hand, the second surface integral vanishes when the field vanishes at the surface. This
form of the energy variation is the same as that given by Bernstein et al. (1958) for
$d\m{S}\cdot\m{B} = 0$. The surface integrals may be relevant, for instance, in the case
of poloidal fields. However, we will not need to worry about these as we will be
considering toroidal fields that vanish at the surface.

\subsection{Type II Superconducting Star}
The magnetic force for a type II superconductor is given by equation (\ref{typeIIforce}),
        \beq
        \m{f}_{\rm mag} = \frac{\m{J}\times\m{B}}{c}
        - \rho \m\nabla\psi_{\rm II} \ ,
        \enq
where $\psi_{\rm II} = \pa F/\pa \rho$, from equation (\ref{pottypeII2}). The current
density is now given through $4\pi\m{J}/c = \m\nabla\times\m{H}$. The magnetic free
energy $F$ is a function of $\rho$ and $B$ and is related to the magnetic field through
equation (\ref{typeIIfree}), $H = 4\pi\pa F/\pa B$. The perturbation of the force gives,
        \beq
        \delta\m{f}_{\rm mag} = \frac{\delta\m{J}\times\m{B}}{c} +
        \frac{\m{J}\times\delta\m{B}}{c} - \delta\rho\m\nabla\psi_{\rm II} -
        \rho\m\nabla\delta\psi_{\rm II} \ .
        \enq
Consider the energy due to the first term of the magnetic force. Following the same
procedure as in the derivation of equation (\ref{stabilitynormal}), we get,
        \beq
        \begin{array}{rcl}\vspace{0.2cm}\displaystyle
        \frac{1}{c}\int \m\xi\cdot\delta\m{J}\times\m{B}\,d V &=& \displaystyle
        - \frac{1}{4\pi}\int \m\xi\times\m{B}\cdot(\m\nabla\times\delta\m{H})\,d V
        \\ &=& \displaystyle
        \frac{1}{4\pi}\oint d\m{S}\cdot\bra{[}\m{B}(\m\xi\cdot\delta\m{H})
        - \m\xi(\m{B}\cdot\delta\m{H})\ket{]} -
        \frac{1}{4\pi}\int \delta\m{H}\cdot\delta\m{B} \, d V
        \end{array}
        \enq
When $B$ vanishes on the surface we can drop the surface integral. On the other hand,
note that we can rewrite the last two terms in the magnetic energy variation as,
        \beq
        \int (\delta\rho\,\m\xi\cdot\m\nabla\psi_{\rm II} +
        \rho\m\xi\cdot\m\nabla\delta\psi_{\rm II})\, d V = \displaystyle
        \int (\delta\rho\,\m\xi\cdot\m\nabla\psi_{\rm II} +
        \delta\rho\,\delta\psi_{\rm II})\, d V
        = \int \delta\rho\Delta\psi_{\rm II} \, d V \ .
        \enq
Here, we have made use of the relation $\Delta = \delta + \m\xi\cdot\m\nabla$, between
Lagrangian and Eulerian perturbations. Thus, the magnetic energy variation for a type II
superconductor becomes, from equation (\ref{energy}),
        \beq
        \begin{array}{rcl}\vspace{0.2cm}\displaystyle
        \delta W_{\rm mag} &=& \displaystyle
        - \frac{1}{2}\int \m\xi\cdot\delta\m{f}_{\rm mag}\,d V
        \\ &=& \displaystyle
        \frac{1}{2}\int \bra{[}\frac{\delta\m{H}\cdot\delta\m{B}}{4\pi}
        - \frac{\m{J}\cdot\delta\m{B}\times\m\xi}{c}
        + \delta\rho\Delta\psi_{\rm II} \ket{]} d V
        + \frac{1}{8\pi}\oint d\m{S}\cdot\bra{[}\m\xi(\m{B}\cdot\delta\m{H})
        - \m{B}(\m\xi\cdot\delta\m{H})\ket{]}
        \end{array}
        \label{stabilitytypeII}
        \enq
This is to be contrasted with the magnetic energy for the normal case given by equation
(\ref{stabilitynormal}). In particular, the first two terms in the volume integrals are
of the same form, with a $\m{B}$ in the normal case replaced by an $\m{H}$ in the
superconducting case. The same is true for the surface integral terms. However, in the
superconducting case there is also an additional term that arises from the potential
$\psi_{\rm II}$, that has no analogue in the normal case.

In the strongly type II superconducting case the magnetic field is a function of density
only, $H = H(\rho)$. On the other hand, in the normal case we have $H = B$. In general,
$H$, $\psi_{\rm II}$ and $F$ will all be functions of $\rho$ and $B$. Using the
definition of the potential $\psi_{\rm II}$ from equation (\ref{pottypeII2}), we get,
        \beq
        \Delta\psi_{\rm II} = \frac{\pa^2 F}{\pa\rho^2}\Delta\rho
        + \frac{\pa^2 F}{\pa\rho\pa B}\Delta B \ .
        \label{stabilitydeltapsiII}
        \enq
We will assume that the form of $\delta\m{B}$ given through equation (\ref{magper}) is
still valid for the superconducting case. Also note the following relations which will be
of use,
        \beq
        \begin{array}{l}\vspace{0.2cm}\displaystyle
        \delta \hB = \frac{\delta\m{B}}{B} - \frac{\delta B \hB}{B}
        \\ \vspace{0.2cm} \displaystyle
        \delta B = \hB\cdot\delta\m{B}
        \\ \vspace{0.2cm} \displaystyle
        \delta\m{H} = \delta H \hB + H\delta\hB
        \\ \displaystyle
        \delta H = \frac{\pa H}{\pa\rho}\delta\rho + \frac{\pa H}{\pa B}\delta B
        \end{array}
        \label{stabilityrel}
        \enq
Note that $\hB \perp \delta\hB$, which also follows from $\delta(\hB\cdot\hB) = 0$. Using
the above relations we have,
        \beq
        \delta\m{H}\cdot\delta\m{B} = \delta H\delta B + \left.\left.
        \frac{H}{B}\right[\delta\m{B}\cdot\delta\m{B} - (\delta B)^2 \right] \ .
        \label{stabilityrel2}
        \enq
Using equation (\ref{typeIIfree}) which relates $H$ and $F$, the perturbation in the
magnetic field can be written as,
        \beq
        \delta H = 4\pi\left(\frac{\pa^2 F}{\pa\rho\pa B}\delta\rho
        + \frac{\pa^2 F}{\pa B^2}\delta B\right)\ .
        \label{stabilitydeltaH}
        \enq
This allows us to express the energy in terms of derivatives of $F$.

For a strongly type II superconductor $H \propto \rho$, equation (\ref{stabilitytypeII})
reduces to (Roberts 1981; Akg\"{u}n 2007),
        \beq
        \begin{array}{rcl}\vspace{0.2cm}\displaystyle
        \delta W_{\rm mag} &=& \displaystyle
        \frac{1}{8\pi}\int \bra{[} \delta\m{H}\cdot\delta\m{B}
        - \delta\m{B}\cdot\m\xi\times(\m\nabla\times\m{H})
        - (\m{H}\cdot\delta\m{B})(\m\nabla\cdot\m\xi)
        + \delta\m{H}\cdot(\m\xi\cdot\m\nabla\m{B})
        - \delta\m{B}\cdot(\m\xi\cdot\m\nabla\m{H}) \ket{]} d V
        \\ && \displaystyle
        + \frac{1}{8\pi}\oint d\m{S}\cdot\bra{[}\m\xi(\delta\m{H}\cdot\m{B}
        + \m{H}\cdot\delta\m{B}) - \m{B}(\m\xi\cdot\delta\m{H})\ket{]}
        \end{array}
        \enq

\subsection{Stability Criteria}\label{sectioncriteria}
Tayler (1973) derives stability conditions for toroidal fields in a normal star in
cylindrical coordinates using the energy principle given by equation
(\ref{stabilitynormal}). The equivalent conditions in spherical coordinates are given by
Goossens \& Veugelen (1978). We will now proceed to derive stability criteria for
toroidal fields in a type II superconducting star, along the same lines. We will take the
magnetic field to be given as a function of density and magnetic induction, $H =
H(\rho,B)$. This will allow us to consider both the strongly type II superconducting case
and the normal case simultaneously. We will closely follow the notation of Goossens \&
Veugelen (1978) in order to facilitate comparisons.

It is clearly sufficient for stability to show that the integrand of the energy of the
perturbations is positive throughout the region of integration,
        \beq
        \delta W = \frac{1}{2}\int {\cal E} d V > 0 \mtext{if}
        {\cal E} > 0 \ .
        \enq
Even if ${\cal E}$ becomes negative in a small region the system is unstable. Define
${\cal E_{\rm p}}$ and ${\cal E_{\rm mag}}$ as the integrands of $\delta W_{\rm p}$ and
$\delta W_{\rm mag}$, i.e. ${\cal E} = {\cal E}_{\rm p} + {\cal E}_{\rm mag}$. As in
previous works (Bernstein et al. 1958; Tayler 1973; Goossens \& Veugelen 1978; and
Roberts 1981) we will drop the gravitational potential perturbation term in ${\cal E_{\rm
p}}$. The hydrostatic and magnetic parts of the energy are then given through equations
(\ref{energy}) and (\ref{stabilitytypeII}), respectively,
        \beq
        \begin{array}{l}\vspace{0.2cm}\displaystyle
        {\cal E_{\rm p}} = \gamma p (\m\nabla\cdot\m\xi)^2 +
        (\m\xi\cdot\m\nabla p) (\m\nabla\cdot\m\xi) - (\m\xi\cdot\m\nabla\phi)
        (\m\nabla\cdot \rho\m\xi) \\ \displaystyle
        {\cal E_{\rm mag}} = \frac{1}{4\pi}\bra{[}\delta\m{H}\cdot\delta\m{B}
        - \delta\m{B}\cdot\m\xi\times(\m\nabla\times\m{H}) \ket{]}
        + \delta\rho\Delta\psi_{\rm II}
        \end{array}
        \label{integrands}
        \enq

The azimuthal angle $\varphi$ does not explicitly appear in any of the coefficients in
these equations, so that we can expand the components of the Lagrangian displacement as,
        \beq
        \xi_r = R(r,\theta)e^{im\varphi} \ , \hspace{0.6cm}
        \xi_\theta = S(r,\theta)e^{im\varphi} \mtext{and} \xi_\phi = i T(r,\theta)e^{im\varphi} \ .
        \enq
Here $m$ is an integer. Since only the real parts are significant, the scalar
multiplications and vector dot products are to be treated as $Z\cdot Z^*$ where $Z^*$
stands for complex conjugate. It will be of great notational convenience to define an
operator $\Lambda$ of a scalar argument $u = u(r,\theta)$,
        \beq
        \Lambda(u) \equiv R\pa_r u + \frac{S\pa_\theta u}{r} \ .
        \label{stabilityop}
        \enq
This is simply the directional derivative along the Lagrangian displacement,
$\m\xi\cdot\m\nabla u = \Lambda(u)e^{im\varphi}$. We will find it convenient to redefine
the $\varphi$ component of the Lagrangian displacement as,
        \beq
        \hat T = \frac{m T}{r\sin\theta} \ .
        \label{stabilityhatT}
        \enq
Also define,
        \beq
        D = \frac{\pa_r(r^2R)}{r^2} + \frac{\pa_\theta(S\sin\theta)}{r\sin\theta}
        - \hat T = D_0 - \hat T \ ,
        \label{stabilitydiv}
        \enq
which is simply the divergence of the Lagrangian displacement, $\m\nabla\cdot\m\xi =
De^{im\varphi}$. Note that $D_0$ is independent of $\hat T$. Using these definitions, we
can express the hydrostatic part given by equation (\ref{integrands}) as,
        \beq
        {\cal E}_{\rm p} = \gamma p D^2
        + \bra{[}\Lambda(p) - \rho\Lambda(\phi)\ket{]} D
        - \Lambda(\rho)\Lambda(\phi) \ .
        \label{stabilityEp}
        \enq

The equations of equilibrium for the unperturbed background state are given by equation
(\ref{eulertypeII}),
        \beq
        \begin{array}{l}\vspace{0.2cm}\displaystyle
        \pa_r p + \rho\pa_r\phi = - \frac{B}{r}\pa_r\left(r\frac{\pa F}{\pa B}\right)
        - \rho\pa_r\left(\frac{\pa F}{\pa\rho}\right)
        \\ \displaystyle
        \pa_\theta p + \rho\pa_\theta\phi =
        - \frac{B}{\sin\theta}\pa_\theta\left(\sin\theta\frac{\pa F}{\pa B}\right)
        - \rho\pa_\theta\left(\frac{\pa F}{\pa\rho}\right)
        \end{array}
        \label{euler2var}
        \enq
Note the notational convention for partial derivatives that we will employ for the
remainder of this section: derivatives with respect to coordinates $x$ will be shortened
as $\pa_x$, while derivatives of the magnetic free energy $F$ with respect to $\rho$ and
$B$ will be explicitly written. Using these equations we can eliminate the pressure
gradient in ${\cal E}_{\rm p}$ and rewrite it in terms of the gravitational and magnetic
forces. Using the definition of the operator $\Lambda$ from equation (\ref{stabilityop}),
we have,
        \beq
        \Lambda(p) = - \rho\Lambda(\phi) - \rho\Lambda\left(\frac{\pa F}{\pa\rho}\right)
        - B\Lambda\left(\frac{\pa F}{\pa B}\right)
        - B\frac{\pa F}{\pa B}\left(\frac{R + S\cot\theta}{r}\right) \ .
        \enq

Next, consider the magnetic part of the integrand given by equation (\ref{integrands}).
Using equation (\ref{stabilityrel2}) for $\delta\m{H}\cdot\delta\m{B}$, we have,
        \beq
        {\cal E_{\rm mag}} = \frac{1}{4\pi}\left[\delta H\delta B
        + \frac{H}{B}\bra{(}|\delta\m{B}|^2 - (\delta B)^2\ket{)}
        - \delta\m{B}\cdot\m\xi\times(\m\nabla\times\m{H}) \right]
        + \delta\rho\Delta\psi_{\rm II} \ .
        \enq
$\Delta\psi_{\rm II}$ and $\delta H$ are given through equations
(\ref{stabilitydeltapsiII}) and (\ref{stabilitydeltaH}), respectively. We can also
express the magnetic field in terms of the free energy through equation
(\ref{typeIIfree}), $H = 4\pi\pa F/\pa B$. The various terms in ${\cal E_{\rm mag}}$ can
be evaluated using the relations given in equation (\ref{stabilityrel}). In particular,
        \beq
        \frac{|\delta\m{B}|^2 - (\delta B)^2}{B^2}
        = \frac{m^2(R^2 + S^2)}{r^2\sin^2\theta}
        \mtext{and}
        \frac{\delta\m{B}\cdot\m\xi\times(\m\nabla\times\m{H})}{H B}
        = \hat X\hat Y + \hat T \hat Y \ ,
        \enq
where we define the following auxiliary quantities,
        \beq
        \hat X = D_0 + \frac{\Lambda(B)}{B} - \frac{R + S \cot\theta}{r}
        \mtext{and}
        \hat Y = \frac{\Lambda(H)}{H} + \frac{R + S\cot\theta}{r} \ .
        \label{aux}
        \enq
The magnetic part can then be written as,
        \beq
        {\cal E_{\rm mag}} = B\frac{\pa F}{\pa B}\left[\frac{m^2(R^2 + S^2)}{r^2\sin^2\theta}
        - \hat X\hat Y - \hat T\hat Y\right]
        + \frac{\pa^2 F}{\pa\rho\pa B}\delta\rho\delta B
        + \frac{\pa^2 F}{\pa B^2}(\delta B)^2
        + \frac{\pa^2 F}{\pa\rho\pa B}\delta\rho\Delta B
        + \frac{\pa^2 F}{\pa\rho^2}\delta\rho\Delta\rho \ ,
        \label{stabilityEmag}
        \enq
where,
        \beq
        \begin{array}{l}\vspace{0.2cm}\displaystyle
        \frac{\delta B}{B} = - \hat X e^{im\varphi}
        \\ \vspace{0.2cm}\displaystyle
        \frac{\Delta B}{B} = - \left[ D_0 - \frac{R + S\cot\theta}{r}\right] e^{im\varphi}
        \\ \vspace{0.2cm}\displaystyle
        \frac{\delta\rho}{\rho} = - \left[D + \frac{\Lambda(\rho)}{\rho}\right] e^{im\varphi}
        \\ \displaystyle
        \frac{\Delta\rho}{\rho} = - D e^{im\varphi}
        \end{array}
        \label{stabilityrel3}
        \enq
We will next consider the $m = 0$ and $m \ne 0$ cases separately.

\subsubsection{The $m = 0$ Case}
In this case $\hat T = 0$ from equation (\ref{stabilityhatT}) and the total energy can be
written as, using equations (\ref{stabilityEp}) and (\ref{stabilityEmag}) for ${\cal
E}_{\rm p}$ and ${\cal E}_{\rm mag}$, respectively,
        \beq
        {\cal E} = {\cal E}_{\rm p} + {\cal E}_{\rm mag}
        = {\cal K}_0 {D_0}^2 + {\cal K}_1 D_0 + {\cal K}_2 \ ,
        \label{stabilitym0}
        \enq
where $D_0$ is defined in equation (\ref{stabilitydiv}). We have, in terms of the
operator $\Lambda$ defined by equation (\ref{stabilityop}),
        \beq
        \begin{array}{l}\vspace{0.2cm}\displaystyle
        {\cal K}_0 = \gamma p + B^2\frac{\pa^2 F}{\pa B^2}
        + 2\rho B\frac{\pa^2 F}{\pa\rho\pa B} + \rho^2\frac{\pa^2 F}{\pa \rho^2}
        \\ \vspace{0.2cm}\displaystyle
        {\cal K}_1 = - 2\rho\,\Lambda(\phi) - 2 \left[ B\frac{\pa F}{\pa B}
        + B^2\frac{\pa^2 F}{\pa B^2} + \rho B \frac{\pa^2 F}{\pa\rho\pa B}
        \right] \frac{R + S\cot\theta}{r}
        \\ \displaystyle
        {\cal K}_2 = - \Lambda(\rho)\Lambda(\phi) - \left[ \Lambda(B)\frac{\pa F}{\pa B}
        + B\Lambda(B)\frac{\pa^2 F}{\pa B^2} + B\Lambda(\rho)\frac{\pa^2 F}{\pa\rho\pa B}
        \right]\frac{R + S\cot\theta}{r}
        + \left[B\frac{\pa F}{\pa B} + B^2\frac{\pa^2 F}{\pa B^2}\right]
        \left(\frac{R + S\cot\theta}{r}\right)^2
        \end{array}
        \label{auxK}
        \enq
All derivatives of $R$ and $S$ are included in $D_0$. By completing the square we get,
        \beq
        {\cal E} = {\cal K}_0 \left(D_0 + \frac{{\cal K}_1}{2{\cal K}_0}\right)^2
        + {\cal K}_2 - \frac{{\cal K}_1^2}{4{\cal K}_0} \ .
        \enq
The first term is non-negative and the remaining terms form a quadratic in $R$ and $S$,
which is also the minimum value of ${\cal E}$ with respect to $D_0$,
        \beq
        {\cal K}_2 - \frac{{\cal K}_1^2}{4{\cal K}_0} = a_0 R^2 + b_0 R S + c_0 S^2 \ .
        \label{quadm0}
        \enq
The subscripts in the coefficients stand for $m = 0$. Define the following auxiliary
quantities,
        \beq
        \begin{array}{l}\vspace{0.2cm}\displaystyle
        U_0 = \frac{1}{r}\left(B\frac{\pa F}{\pa B}
        + B^2\frac{\pa^2 F}{\pa B^2} + \rho B \frac{\pa^2 F}{\pa\rho\pa B}\right)
        \\ \vspace{0.2cm}\displaystyle
        U_1 = \frac{1}{r}\left(\pa_r B\frac{\pa F}{\pa B}
        + B\pa_r B\frac{\pa^2 F}{\pa B^2}
        + B\pa_r\rho\frac{\pa^2 F}{\pa\rho\pa B}\right)
        \\ \vspace{0.2cm}\displaystyle
        U_2 = \frac{1}{r^2}\left(\pa_\theta B\frac{\pa F}{\pa B}
        + B\pa_\theta B\frac{\pa^2 F}{\pa B^2}
        + B\pa_\theta\rho\frac{\pa^2 F}{\pa\rho\pa B}\right)
        \\ \displaystyle
        U_3 = \frac{1}{r^2}\left(B\frac{\pa F}{\pa B}
        + B^2\frac{\pa^2 F}{\pa B^2}\right)
        \end{array}
        \label{auxU}
        \enq
We then find that the coefficients in the quadratic are given by,
        \beq
        \begin{array}{lcl}\vspace{0.2cm}\displaystyle
        a_0 &=& \displaystyle - \pa_r\rho\,\pa_r\phi - U_1 + U_3
        - \left.\left.\frac{1}{{\cal K}_0}\right(\rho\pa_r\phi + U_0\right)^2
        \\ \vspace{0.2cm}\displaystyle
        b_0 &=& \displaystyle - \frac{\pa_r\rho\,\pa_\theta\phi}{r}
        - \frac{\pa_\theta\rho\,\pa_r\phi}{r}
        - U_1\cot\theta - U_2 + 2U_3\cot\theta
        - \left.\left.\frac{2}{{\cal K}_0}\right(\rho\pa_r\phi + U_0\right)
        \left(\frac{\rho\pa_\theta\phi}{r} + U_0\cot\theta\right)
        \\ \displaystyle
        c_0 &=& \displaystyle - \frac{\pa_\theta\rho\,\pa_\theta\phi}{r^2} - U_2\cot\theta
        + U_3\cot^2\theta - \frac{1}{{\cal K}_0}
        \left(\frac{\rho\pa_\theta\phi}{r} + U_0\cot\theta\right)^2
        \end{array}
        \label{stabilitycoefm0}
        \enq
A sufficient condition for stability is that the quadratic form be always positive
throughout the integration region. This corresponds to the following conditions, which
are not all independent,
        \beq
        a > 0 \ , \hspace{0.2cm} c > 0 \mtext{and} b^2 < 4 ac \ .
        \label{con}
        \enq
When these conditions are satisfied the star is stable, therefore these are
\emph{sufficient} conditions for stability. If we can show that the star is unstable as
soon as one of these conditions is violated, then we will have shown that the conditions
are also \emph{necessary} for stability. For the $m = 0$ case it can be shown that the
interchange instability sets in when these conditions fail, as will be proven in a later
section. Therefore, these conditions are necessary and sufficient conditions for the $m =
0$ case. However, the same will not be true in general for the $m \ne 0$ case, as will be
discussed later.

One way of deriving these conditions is to consider the minimum value of the quadratic
form $Q = a R^2 + b R S + c S^2$ with respect to $S$ (or equivalently, $R$). For a
minimum we need $dQ/dS = 0$ and $d^2Q/dS^2 > 0$. Substituting the value of $S$ that
minimizes $Q$ and requiring that $Q > 0$ we get the condition $b^2 < 4 ac$, while the
second requirement gives $c > 0$. These two conditions then imply the third, $a > 0$.

We can now consider special cases. In the strongly type II superconducting case the
magnetic field is a function of density, $H = H(\rho)$ and the magnetic free energy is
given by equation (\ref{typeIIfree}) as $F = HB/4\pi$. In particular, consider a power
law of the form $H \propto \rho^\sigma$. From equations (\ref{auxK}) and (\ref{auxU}), we
have,
        \beq
        {\cal K}_0 = \gamma p + \frac{\sigma(\sigma + 1)HB}{4\pi} \ , \hspace{0.2cm}
        U_0 = \frac{(\sigma + 1)HB}{4\pi r} \ , \hspace{0.2cm}
        U_1 = \frac{\pa_r(HB)}{4\pi r} \ , \hspace{0.2cm}
        U_2 = \frac{\pa_\theta(HB)}{4\pi r^2} \mtext{and}
        U_3 = \frac{HB}{4\pi r^2} \ ,
        \enq
so that the coefficients become,
        \beq
        \begin{array}{lcl}\vspace{0.2cm}
        a_0 &=& \displaystyle - \pa_r\rho\,\pa_r\phi - \frac{\pa_r(HB)}{4\pi r}
        + \frac{HB}{4\pi r^2}
        - \frac{1}{{\cal K}_0}\left(\rho\pa_r\phi
        + \frac{(\sigma + 1)HB}{4\pi r}\right)^2
        \\ \vspace{0.2cm}
        b_0 &=& \displaystyle - \frac{\pa_r\rho\,\pa_\theta\phi}{r}
        - \frac{\pa_\theta\rho\,\pa_r\phi}{r} - \frac{\pa_r(HB)}{4\pi r}\cot\theta
        - \frac{\pa_\theta(HB)}{4\pi r^2} + \frac{HB}{2\pi r^2}\cot\theta
        \\ \vspace{0.2cm}
        && \displaystyle - \frac{2}{r{\cal K}_0}
        \left(\rho\pa_r\phi + \frac{(\sigma + 1)HB}{4\pi r}\right)
        \left(\rho\pa_\theta\phi + \frac{(\sigma + 1)HB}{4\pi}\cot\theta\right)
        \\
        c_0 &=& \displaystyle - \frac{\pa_\theta\rho\,\pa_\theta\phi}{r^2}
        - \frac{\pa_\theta(HB)}{4\pi r^2}\cot\theta
        + \frac{HB}{4\pi r^2}\cot^2\theta - \frac{1}{r^2{\cal K}_0}
        \left(\rho\pa_\theta\phi + \frac{(\sigma + 1)HB}{4\pi}\cot\theta\right)^2
        \end{array}
        \label{stabilityHrho}
        \enq

On the other hand, in the normal conducting case the magnetic field and induction are
equal $H = B$, and the free energy is $F = B^2/8\pi$, so that from equations (\ref{auxK})
and (\ref{auxU}), we have,
        \beq
        {\cal K}_0 = \gamma p + \frac{B^2}{4\pi} \ , \hspace{0.2cm}
        U_0 = \frac{B^2}{2\pi r} \ , \hspace{0.2cm}
        U_1 = \frac{B\pa_r B}{2\pi r} \ , \hspace{0.2cm}
        U_2 = \frac{B\pa_\theta B}{2\pi r^2} \mtext{and}
        U_3 = \frac{B^2}{2\pi r^2} \ ,
        \enq
and the coefficients are given by,
        \beq
        \begin{array}{lcl}\vspace{0.2cm}\displaystyle
        a_0 &=& \displaystyle - \pa_r\rho\,\pa_r\phi - \frac{B\pa_r B}{2\pi r}
        + \frac{B^2}{2\pi r^2}
        - \frac{1}{{\cal K}_0}\left(\rho\pa_r\phi + \frac{B^2}{2\pi r}\right)^2
        \\ \vspace{0.2cm}\displaystyle
        b_0 &=& \displaystyle - \frac{\pa_r\rho\,\pa_\theta\phi}{r}
        - \frac{\pa_\theta\rho\,\pa_r\phi}{r} - \frac{B\pa_r B}{2\pi r}\cot\theta
        - \frac{B\pa_\theta B}{2\pi r^2} + \frac{B^2}{\pi r^2}\cot\theta
        \\ \vspace{0.2cm} && \displaystyle
        - \frac{2}{r{\cal K}_0}\left(\rho\pa_r\phi + \frac{B^2}{2\pi r}\right)
        \left(\rho\pa_\theta\phi + \frac{B^2}{2\pi}\cot\theta\right)
        \\ \displaystyle
        c_0 &=& \displaystyle - \frac{\pa_\theta\rho\,\pa_\theta\phi}{r^2}
        - \frac{B\pa_\theta B}{2\pi r^2}\cot\theta + \frac{B^2}{2\pi r^2}\cot^2\theta
        - \frac{1}{r^2{\cal K}_0}
        \left(\rho\pa_\theta\phi + \frac{B^2}{2\pi}\cot\theta\right)^2
        \end{array}
        \label{stabilitynormalm0}
        \enq
These are the same as the results given by Goossens \& Veugelen (1978).\footnote{Note
that there is a typo in equation (13) of Goossens \& Veugelen (1978).}

\subsubsection{The $m \ne 0$ Case}
When $m \ne 0$, the hydrostatic and magnetic parts of the energy are given by equations
(\ref{stabilityEp}) and (\ref{stabilityEmag}), respectively. In this case, the integrand
${\cal E} = {\cal E_{\rm p}} + {\cal E_{\rm mag}}$ is quadratic in the rescaled $\varphi$
component of the Lagrangian displacement $\hat T$, defined by equation
(\ref{stabilityhatT}), and does not contain any derivatives of it. Therefore, we can
write the energy as,
        \beq
        {\cal E}
        = {\cal E}_o + \alpha \hat T^2 + \beta \hat T
        + B\frac{\pa F}{\pa B}\left[\frac{m^2(R^2 + S^2)}{r^2\sin^2\theta}\right] \ ,
        \enq
where ${\cal E}_o$ is the energy for the $m = 0$ case, given by equation
(\ref{stabilitym0}), and we define,
        \beq
        \begin{array}{l}\vspace{0.2cm}\displaystyle
        \alpha = \gamma p + \rho^2\frac{\pa^2 F}{\pa\rho^2}
        \\ \displaystyle
        \beta = - 2\left(\gamma p + \rho B\frac{\pa^2 F}{\pa\rho\pa B}
        + \rho^2\frac{\pa^2 F}{\pa\rho^2}\right)D_0
        + 2\rho B\frac{\pa^2 F}{\pa\rho\pa B}\left(\frac{R + S\cot\theta}{r}\right)
        + 2\rho\Lambda(\phi)
        \end{array}
        \label{auxalpha}
        \enq
${\cal E}_o$ is independent of $\hat T$. We therefore have $d^2 {\cal E} /d\hat T^2 =
2\alpha$. The $\gamma p$ term in $\alpha$ will be the dominant term for the cases of
interest to us, so that $d^2 {\cal E} /d\hat T^2 > 0$, and consequently ${\cal E}$ can be
minimized with respect to $\hat T$. Setting $d{\cal E}/d\hat T = 0$ we get the value that
minimizes the energy, $\hat T = - \beta/2\alpha$. Substituting this back into the energy
we find the minimum as,
        \beq
        {\cal E} = {\cal E}_o - \frac{\beta^2}{4\alpha}
        + B\frac{\pa F}{\pa B}\left[\frac{m^2(R^2 + S^2)}{r^2\sin^2\theta}\right] \ .
        \enq
As was done in equation (\ref{stabilitym0}) for the $m = 0$ case, we can once again group
together terms of different order in $D_0$, defined by equation (\ref{stabilitydiv}),
        \beq
        {\cal E} = {\cal L}_0{D_0}^2 + {\cal L}_1 D_0 + {\cal L}_2 \ .
        \enq
For notational convenience, define a set of auxiliary quantities,
        \beq
        \begin{array}{l}\vspace{0.2cm}\displaystyle
        V_0 = \alpha^{-1/2}\left(\gamma p + \rho B\frac{\pa^2 F}{\pa\rho\pa B}
        + \rho^2\frac{\pa^2 F}{\pa\rho^2}\right)
        \\ \vspace{0.2cm}\displaystyle
        V_1 = \alpha^{-1/2}\left(\rho\pa_r\phi
        + \frac{\rho B}{r}\frac{\pa^2 F}{\pa\rho\pa B}\right)
        \\ \displaystyle
        V_2 = \alpha^{-1/2}\left(\frac{\rho\pa_\theta\phi}{r}
        + \frac{\rho B\cot\theta}{r}\frac{\pa^2 F}{\pa\rho\pa B}\right)
        \end{array}
        \label{auxV}
        \enq
and,
        \beq
        \begin{array}{l}\vspace{0.2cm}\displaystyle
        W_1 = - \rho\pa_r\phi - U_0 + V_0V_1
        \\ \displaystyle
        W_2 = - \frac{\rho\pa_\theta\phi}{r} - U_0\cot\theta + V_0V_2
        \end{array}
        \label{auxW}
        \enq
where $\alpha$ is defined in equation (\ref{auxalpha}), and $U_0$ is defined in equation
(\ref{auxU}). Also invoking the definitions of ${\cal K}_i$ from equation (\ref{auxK}),
we have,
        \beq
        \begin{array}{l}\vspace{0.2cm}\displaystyle
        {\cal L}_0 = {\cal K}_0 - {V_0}^2 = B^2\frac{\pa^2 F}{\pa B^2}
        - \frac{1}{\alpha}\left(\rho B\frac{\pa^2 F}{\pa\rho\pa B}\right)^2
        \\ \vspace{0.2cm}\displaystyle
        {\cal L}_1 = {\cal K}_1 + 2V_0(V_1 R + V_2 S) = 2(W_1R + W_2S)
        \\ \displaystyle
        {\cal L}_2 = {\cal K}_2 - (V_1 R + V_2 S)^2
        + B\frac{\pa F}{\pa B}\left[\frac{m^2(R^2 + S^2)}{r^2\sin^2\theta}\right]
        \end{array}
        \label{auxL}
        \enq
Rearranging the terms we get,
        \beq
        {\cal E} = {\cal L}_0 \left(D_0 + \frac{{\cal L}_1}{2{\cal L}_0}\right)^2
        + {\cal L}_2 - \frac{{\cal L}_1^2}{4{\cal L}_0} \ .
        \label{stabilitym1}
        \enq
Note that ${\cal L}_0$ is not necessarily positive, so unlike in the $m = 0$ case, it is
not obvious that the first term is positive definite. In fact, for the strongly type II
case where the free energy is of the form $F = H(\rho)B/4\pi$, we have ${\cal L}_0 < 0$.
On the other hand, for the normal case $F = B^2/8\pi$, so that ${\cal L}_0 > 0$. For
negative ${\cal L}_0$ the system is unstable since we can find displacement fields with
sufficiently large derivatives which will make the $D_0$ term dominant in the energy.
Therefore, for stability we must require ${\cal L}_0 > 0$, or using the definitions of
equation (\ref{auxL}),
        \beq
        B^2\frac{\pa^2 F}{\pa B^2} >
        \left(\rho B\frac{\pa^2 F}{\pa\rho\pa B}\right)^2
        \left/\left(\gamma p + \rho^2\frac{\pa^2 F}{\pa\rho^2}\right)\right. \ .
        \label{MPRL0}
        \enq
This is a necessary but not sufficient condition for stability. This is related to what
we will refer to as the Muzikar--Pethick--Roberts (MPR) instability (Muzikar \& Pethick
1981; Roberts 1981), which we will discuss in more detail in a later section.

Another way of looking at equation (\ref{stabilitym1}) is to say that when ${\cal L}_0 >
0$, the energy can be minimized with respect to $D_0$. The minimum is a quadratic in $R$
and $S$, just like equation (\ref{quadm0}) for the $m = 0$ case,
        \beq
        {\cal L}_2 - \frac{{\cal L}_1^2}{4{\cal L}_0} = a_m R^2 + b_m R S + c_m S^2 \ .
        \label{quad}
        \enq
The coefficients are given as, using the definitions of $U_i$, $V_i$ and $W_i$ made in
equations (\ref{auxU}), (\ref{auxV}) and (\ref{auxW}),
        \beq
        \begin{array}{l}\vspace{0.2cm}\displaystyle
        a_m = - \pa_r\rho\,\pa_r\phi - U_1 + U_3 - {V_1}^2
        + \frac{m^2 B}{r^2\sin^2\theta}\frac{\pa F}{\pa B}
        - \frac{W_1^2}{{\cal L}_0}
        \\ \vspace{0.2cm}\displaystyle
        b_m = - \frac{\pa_r\rho\,\pa_\theta\phi}{r}
        - \frac{\pa_\theta\rho\,\pa_r\phi}{r} - U_1\cot\theta - U_2 + 2U_3\cot\theta
        - 2 V_1V_2 - \frac{2W_1W_2}{{\cal L}_0}
        \\ \displaystyle
        c_m = - \frac{\pa_\theta\rho\,\pa_\theta\phi}{r^2} - U_2\cot\theta
        + U_3\cot^2\theta - {V_2}^2
        + \frac{m^2 B}{r^2\sin^2\theta}\frac{\pa F}{\pa B}
        - \frac{W_2^2}{{\cal L}_0}
        \end{array}
        \label{stabilitycoefm1}
        \enq
This quadratic is positive if the coefficients satisfy the conditions listed in equation
(\ref{con}). However, the system will be definitely stable only when ${\cal L}_0 > 0$. On
the other hand, if these conditions are violated, i.e. if the quadratic is negative, then
the system is unstable regardless of the sign of ${\cal L}_0$. Also note that, clearly,
the $|m| = 1$ case is the worst instability, as noted previously for the normal case by
Tayler (1973) and Goossens \& Veugelen (1978). On the other hand, when ${\cal L}_0 < 0$
the energy is maximized with respect to $D_0$, and it is always possible to find a
Lagrangian displacement field with sufficiently large derivatives that will make the
system unstable.

The coefficients for the strongly type II case can be obtained by setting $F = HB/4\pi$.
On the other hand, for the normal case we have $F=B^2/8\pi$, and the coefficients reduce
to,
        \beq
        \begin{array}{l}\vspace{0.2cm}\displaystyle
        a_m = - \pa_r\rho\,\pa_r\phi - \frac{(\rho\pa_r\phi)^2}{\gamma p}
        - \frac{B\pa_r B}{2\pi r} - \frac{B^2}{2\pi r^2}
        + \frac{m^2 B^2}{4\pi r^2\sin^2\theta}
        \\ \vspace{0.2cm}\displaystyle
        b_m = - \frac{\pa_r\rho\,\pa_\theta\phi}{r}
        - \frac{\pa_\theta\rho\,\pa_r\phi}{r}
        - \frac{2\rho^2\pa_r\phi\,\pa_\theta\phi}{\gamma pr}
        - \frac{B\pa_r B}{2\pi r}\cot\theta - \frac{B\pa_\theta B}{2\pi r^2}
        - \frac{B^2}{\pi r^2}\cot\theta
        \\ \displaystyle
        c_m = - \frac{\pa_\theta\rho\,\pa_\theta\phi}{r^2}
        - \frac{(\rho\pa_\theta\phi)^2}{\gamma p r^2}
        - \frac{B\pa_\theta B}{2\pi r^2}\cot\theta
        - \frac{B^2}{2\pi r^2}\cot^2\theta + \frac{m^2B^2}{4\pi r^2\sin^2\theta}
        \end{array}
        \label{stabilitynormalm1}
        \enq
These are the same as the results given by Goossens \& Veugelen (1978).

In the next two sections we will consider the special cases of the completely normal
conducting star and the strongly type II superconducting star with $H\propto\rho$. The
coefficients $a$, $b$ and $c$ (given by equations \ref{stabilitycoefm0} and
\ref{stabilitycoefm1}) have hydrostatic terms that are of the form $\pa_r\rho$ and
$\pa_\theta\rho$, and magnetic terms of the order of the magnetic free energy $F$. The
radial dependence of the background quantities arises from the much stronger hydrostatic
forces, while the $\theta$ dependence arises as a result of magnetic forces. Therefore,
$\pa_\theta\rho \sim F \ll \pa_r\rho$. We will calculate the coefficients to first order
in the magnetic energy, which is much smaller than the hydrostatic terms. We will assume
that the perturbations and the background state have the same index, thus neglecting
buoyancy effects. If we include buoyancy, then to leading order, the coefficient $a$ will
be a buoyant term, $c$ will be a purely magnetic term, and $b$ will be the product of a
buoyant term and a magnetic term. Thus, $b^2 \ll 4ac$, and the stability conditions
(given by equation \ref{con}) will reduce to $a > 0$ and $c > 0$. The first condition is
necessary for stability to buoyancy, and the second is the same condition on the magnetic
field as without buoyancy. We will consider the effects of multi-fluid composition in
more detail in future work.

\subsection{Stability Criteria for a Normal Star} \label{sectioncritnormal}
We will now examine the stability of a particular magnetic field
configuration in a normal star. The equilibrium equations in this case are, from equation
(\ref{euler2var}),
        \beq
        \begin{array}{l}\vspace{0.2cm}\displaystyle
        \pa_r p + \rho\pa_r\phi = - \frac{B\pa_r(B r)}{4\pi r}
        \\ \displaystyle
        \pa_\theta p + \rho\pa_\theta\phi = -
        \frac{B\pa_\theta(B\sin\theta)}{4\pi\sin\theta}
        \end{array}
        \enq

Let $p_o$, $\rho_o$ and $\phi_o$ refer to the hydrostatic equilibrium in the absence of
magnetic fields. This equilibrium is spherically symmetric and is simply given through,
        \beq
        \pa_r p_o + \rho_o\pa_r\phi_o = 0 \ .
        \label{eulernomag}
        \enq
The difference between $p_o$, $\rho_o$ and $\phi_o$ and the corresponding quantities $p$,
$\rho$ and $\phi$ in the presence of magnetic fields is of the order of the magnetic
pressure $\sim B^2$, which we assume to be small compared to the hydrostatic pressure.
Therefore, using the equations of equilibrium we can expand equation
(\ref{stabilitynormalm0}) for $m = 0$ to lowest order in $B^2$,
        \beq
        \begin{array}{l}\vspace{0.2cm}\displaystyle
        a_0 \approx \frac{B^2}{4\pi r^2}\left(\frac{d\ln\rho_o}{d\ln r}\right)^2 +
        \left[\frac{3B^2}{4\pi r^2} - \frac{B\pa_r B}{4\pi r}\right]
        \frac{d\ln\rho_o}{d\ln r} - \frac{B\pa_r B}{2\pi r} + \frac{B^2}{2\pi r^2}
        \\ \vspace{0.2cm}\displaystyle
        b_0 \approx \left[\frac{3B^2}{4\pi r^2}\cot\theta
        - \frac{B\pa_\theta B}{4\pi r^2}\right]\frac{d\ln\rho_o}{d\ln r}
        - \frac{B\pa_r B}{2\pi r}\cot\theta - \frac{B\pa_\theta B}{2\pi r^2}
        + \frac{B^2}{\pi r^2}\cot\theta
        \\ \displaystyle
        c_0 \approx - \frac{B\pa_\theta B}{2\pi r^2}\cot\theta
        + \frac{B^2}{2\pi r^2}\cot^2\theta
        \end{array}
        \label{normalm0appr}
        \enq
On the other hand, for $m = 1$, we have, from equation (\ref{stabilitynormalm1}),
        \beq
        \begin{array}{l}\vspace{0.2cm}\displaystyle
        a_m \approx - \left(2 + \frac{d\ln\rho_o}{d\ln r}\right)
        \left(\frac{B^2}{4\pi r^2} + \frac{B\pa_r B}{4\pi r}\right)
        + \frac{B^2}{4\pi r^2\sin^2\theta}
        \\ \vspace{0.2cm}\displaystyle
        b_m \approx - \left(2 + \frac{d\ln\rho_o}{d\ln r}\right)
        \left(\frac{B^2}{4\pi r^2}\cot\theta + \frac{B\pa_\theta B}{4\pi r^2}\right)
        - \frac{B\pa_r B}{2\pi r}\cot\theta
        - \frac{B^2}{2\pi r^2}\cot\theta
        \\ \displaystyle
        c_m \approx - \frac{B\pa_\theta B}{2\pi r^2}\cot\theta
        - \frac{B^2}{2\pi r^2}\cot^2\theta + \frac{B^2}{4\pi r^2\sin^2\theta}
        \end{array}
        \label{normalm1appr}
        \enq

We will now consider a specific example. Let the equation of state be given by a $\gamma
= 2$ polytrope, where the background density profile is $\rho = \rho_c\sin x/x$, in terms
of the dimensionless radial coordinate $x = r/r_o$. Assume a magnetic field of the form
given by equation (\ref{toroidal-normalB}),
        \beq
        B(r,\theta) = \hat B_o\left(\frac{\rho}{\rho_c}\right)^{(n+2)/4}
        \left(\frac{r}{r_o}\right)^{n/2}\sin^{n/2}\theta
        = \hat B_o x^{(n-2)/4}\sin^{(n+2)/4}x\sin^{n/2}\theta \ ,
        \enq
where $n \geqslant 1$. Then, for $m = 0$, the coefficients become, from equation
(\ref{normalm0appr}),
        \beq
        \begin{array}{l}\vspace{0.2cm}\displaystyle
        a_0 \approx \frac{\hat{B}_o^2}{16\pi r_o^2}(2 - n)
        (1 + x\cot x)^2 x^{(n-6)/2}\sin^{(n+2)/2} x\sin^n\theta
        \\ \vspace{0.2cm}\displaystyle
        b_0 \approx \frac{\hat{B}_o^2}{4\pi r_o^2}(2 - n)
        (1 + x\cot x) x^{(n-6)/2}\sin^{(n+2)/2} x\sin^{n-1}\theta\cos\theta
        \\ \displaystyle
        c_0 \approx \frac{\hat{B}_o^2}{4\pi r_o^2}(2 - n)
        x^{(n-6)/2}\sin^{(n+2)/2} x \sin^{n-2}\theta\cos^2\theta
        \end{array}
        \enq
Note that $b_0\!^2 = 4a_0c_0$, so that the quadratic forms a complete square, i.e.
$a_0R^2 + b_0RS + c_0S^2 = a_0 (R + b_0S/2a_0)^2$. However, for $n > 2$, we have $a_0 <
0$ and $c_0 < 0$, and the conditions for stability (equation \ref{con}) are violated.
Thus, only fields with $1 \leqslant n \leqslant 2$ are marginally stable for $m = 0$.

On the other hand, for $m = 1$, we have, from equation (\ref{normalm1appr}),
        \beq
        \begin{array}{l}\vspace{0.2cm}\displaystyle
        a_m \approx \frac{\hat B_o^2}{16\pi r_o^2}
        \left[4 - (n+2)(1 + x\cot x)^2\sin^2\theta\right]
        x^{(n-6)/2}\sin^{(n+2)/2}x\sin^{n-2}\theta
        \\ \vspace{0.2cm}\displaystyle
        b_m \approx - \frac{\hat B_o^2}{4\pi r_o^2} (n+2)
        (1 + x\cot x)x^{(n-6)/2}\sin^{(n+2)/2}x\sin^{n-1}\theta\cos\theta
        \\ \displaystyle
        c_m \approx \frac{\hat B_o^2}{4\pi r_o^2}
        \left[1 - (n+2)\cos^2\theta\right]
        x^{(n-6)/2}\sin^{(n+2)/2}x\sin^{n-2}\theta
        \end{array}
        \enq
Since $a_m$ and $c_m$ become negative in some regions, they violate the stability
conditions given by equation (\ref{con}). Consequently, the normal magnetic field is
unstable for $m = 1$. Thus, we might expect $n = 1$ models with both normal and
superconducting regions to be unstable. Poloidal fields may stabilize the star, as in
normal conductors (Tayler 1973; Wright 1973; Braithwaite \& Nordlund 2006), and we
consider adding them in a following section.

\subsection{Stability Criteria for a Superconducting Star with $H\propto\rho$}
\label{sectioncrittypeII}

We will now consider the strongly type II superconducting case with $H\propto\rho$ (i.e.
$\sigma = 1$) in more detail. In this case $F = H B/4\pi$, and the equations of
equilibrium (equation \ref{euler2var}) explicitly give,
        \beq
        \begin{array}{l}\vspace{0.2cm}\displaystyle
        \pa_r p + \rho\pa_r\phi = - \frac{B\pa_r(H r)}{4\pi r} - \frac{H\pa_r B}{4\pi}
        \\ \displaystyle
        \pa_\theta p + \rho\pa_\theta\phi = -
        \frac{H B}{4\pi}\cot\theta - \frac{H\pa_\theta B}{4\pi}
        \end{array}
        \enq
Using these equations as well as the equation of equilibrium in the absence of magnetic
fields (equation \ref{eulernomag}), we can expand the coefficients for $m = 0$ (equation
\ref{stabilityHrho}) to lowest order in $HB$,
        \beq
        \begin{array}{l}\vspace{0.2cm}\displaystyle
        a_0 \approx \frac{HB}{2\pi r^2}\left(\frac{d\ln\rho_o}{d\ln r}\right)^2 +
        \left[\frac{3HB}{4\pi r^2} - \frac{\pa_r(HB)}{4\pi r}\right]
        \frac{d\ln\rho_o}{d\ln r} - \frac{\pa_r(HB)}{4\pi r} + \frac{HB}{4\pi r^2}
        \\ \vspace{0.2cm}\displaystyle
        b_0 \approx \left[\frac{3HB}{4\pi r^2}\cot\theta
        - \frac{\pa_\theta(HB)}{4\pi r^2}\right]\frac{d\ln\rho_o}{d\ln r}
        - \frac{\pa_r(HB)}{4\pi r}\cot\theta - \frac{\pa_\theta(HB)}{4\pi r^2}
        + \frac{HB}{2\pi r^2}\cot\theta
        \\ \displaystyle
        c_0 \approx - \frac{\pa_\theta(HB)}{4\pi r^2}\cot\theta
        + \frac{HB}{4\pi r^2}\cot^2\theta
        \end{array}
        \label{typeIIm0appr}
        \enq

For a $\gamma = 2$ polytrope we have $\rho = \rho_c\sin x/x$. Consider a magnetic field
of the form given by equation (\ref{toroidal-typeIIB}), for $\sigma = 1$,
        \beq
        B(r,\theta) = B_o\left(\frac{\rho}{\rho_c}\right)^n
        \left(\frac{r}{r_o}\right)^n\sin^n\theta
        = B_o\sin^n x\sin^n\theta \ ,
        \enq
where $n \geqslant 1$. In particular, we get, from equation (\ref{typeIIm0appr}),
        \beq
        c_0 \approx \frac{H_cB_o}{4\pi r_o^2}(1 - n)
        x^{- 3}\sin^{n+1} x\sin^{n-2}\theta\cos^2\theta \ .
        \enq
For all $n > 1$ this is negative, thus immediately violating one of the conditions for
stability (equation \ref{con}). For $n = 1$ all three coefficients vanish to lowest order
in $HB$, implying that the magnetic field is marginally stable. In Appendix A, we show
that this result is true for any $H(\rho,B)$.

For $m \ne 0$, we have (equation \ref{auxL}),
        \beq
        {\cal L}_0 = - \frac{1}{\gamma p}\left(\frac{HB}{4\pi}\right)^2 < 0 \ ,
        \enq
which implies that even if the conditions given in equation (\ref{con}) are met the
system will still be unstable. This is the MPR instability and will be discussed in a
following section in more detail.

\subsection{Interchange Instability}
In this section we will show that the $m = 0$ stability conditions correspond to the
stability criteria for the interchange of two magnetic flux tubes, as demonstrated for
the normal case by Tayler (1973). Consider two axisymmetric flux tubes located at
coordinates $r$, $\theta$ and at $r + \delta r$, $\theta + \delta\theta$, and having
volumes $V$ and $V + \delta V$ and corresponding cross-sections $A$ and $A+\delta A$,
respectively. We will assume that the interchange is adiabatic so that the mass $\rho V$,
magnetic flux $BA$ and $pV^\gamma$ are all conserved.

Let the pressure, density and magnetic induction of the two tubes initially be,
        \beq
        \begin{array}{lccc}\vspace{0.2cm}
        \mbox{at $r$, $\theta$:} & p & \rho & B
        \\
        \mbox{at $r+\delta r$, $\theta+\delta\theta$:}
        & p+\delta p & \rho+\delta\rho & B+\delta B
        \end{array}
        \label{interpre}
        \enq
After the interchange the corresponding quantities are, defining a cylindrical radius by
$\varpi = r\sin\theta$,
        \beq
        \begin{array}{lccc}\vspace{0.2cm}
        \mbox{at $r$, $\theta$:}
        & \displaystyle \frac{(p+\delta p)(V+\delta V)^\gamma}{V^\gamma}
        & \displaystyle \frac{(\rho+\delta\rho)(V+\delta V)}{V}
        & \displaystyle \frac{(B+\delta B)(V+\delta V)\varpi}
        {V(\varpi+\delta\varpi)}
        \\
        \mbox{at $r+\delta r$, $\theta+\delta\theta$:}
        & \displaystyle \frac{pV^\gamma}{(V+\delta V)^\gamma}
        & \displaystyle \frac{\rho V}{V+\delta V}
        & \displaystyle \frac{BV(\varpi+\delta\varpi)}{(V+\delta V)\varpi}
        \end{array}
        \label{interpost}
        \enq

The total energy is the sum of internal, magnetic and gravitational energies. Without
loss of generality, we can take the zero of the gravitational potential to be at $r$,
$\theta$. Prior to the interchange, the energy is,
        \beq
        E_i = \frac{pV}{\gamma - 1}
        + \frac{(p+\delta p)(V+\delta V)}{\gamma - 1}
        + F(\rho,B)V + F(\rho+\delta\rho,B+\delta B)(V+\delta V)
        + (\rho+\delta\rho)(V+\delta V)\delta\phi \ .
        \enq
Here $F$ is the magnetic free energy. After the interchange, we have,
        \beq
        E_f = \frac{(p+\delta p)(V+\delta V)^\gamma}{(\gamma-1)V^{\gamma-1}}
        + \frac{pV^\gamma}{(\gamma - 1)(V+\delta V)^{\gamma-1}}
        + F(\rho_1,B_1)V + F(\rho_2,B_2)(V+\delta V) + \rho V\delta\phi \ .
        \label{Ef}
        \enq
Here $\rho_1$ and $B_1$ are the new density and induction at $r$, $\theta$, and $\rho_2$
and $B_2$ are the corresponding quantities at $r+\delta r$, $\theta+\delta\theta$
(equation \ref{interpost}). We need to calculate the energy difference resulting from the
interchange to second order,
        \beq
        \Delta E = E_f - E_i = \Delta E_p + \Delta E_m \ .
        \enq
Here for notational convenience we denote by $\Delta E_p$ the change in the internal and
gravitational energies, and $\Delta E_m$ is the change in the magnetic energy. To second
order we have,
        \beq
        \Delta E_p \approx \gamma p \frac{(\delta V)^2}{V}
        + (\delta p - \rho\delta\phi)\delta V - V\delta\rho\delta\phi \ .
        \label{deltaEp}
        \enq
Using the equations of equilibrium (equation \ref{euler2var}) we have,
        \beq
        \delta p = - \rho\delta\phi
        - B\frac{\pa F}{\pa B}\frac{\delta\varpi}{\varpi}
        - B\frac{\pa^2 F}{\pa\rho\pa B}\delta\rho
        - \rho\frac{\pa^2 F}{\pa\rho\pa B}\delta B
        - B\frac{\pa ^2 F}{\pa B^2}\delta B
        - \rho\frac{\pa ^2 F}{\pa\rho^2}\delta\rho \ .
        \label{deltap}
        \enq

The magnetic term in the energy change is lengthy. First, note that,
        \beq
        F(\rho+\delta\rho,B+\delta B) \approx F(\rho,B) + \frac{\pa F}{\pa\rho}\delta\rho
        + \frac{\pa F}{\pa B}\delta B + \frac{1}{2}\frac{\pa^2 F}{\pa\rho^2}(\delta\rho)^2
        + \frac{\pa^2 F}{\pa\rho\pa B}\delta\rho\delta B
        + \frac{1}{2}\frac{\pa^2 F}{\pa B^2}(\delta B)^2 \ .
        \enq
We can write the magnetic terms in $E_f$ (equation \ref{Ef}) as $F(\rho_i,B_i) = F(\rho +
\delta\rho_i,B+\delta B_i)$, so they can be expanded in a similar fashion. Here we have,
to second order,
        \beq
        \begin{array}{rcl}\vspace{0.2cm}
        \delta\rho_1 \!\!&=&\!\! \displaystyle
        \rho_1 - \rho = \frac{(\rho+\delta\rho)(V+\delta V)}{V} - \rho
        = \rho\left[\frac{\delta\rho}{\rho} + \frac{\delta V}{V}
        + \frac{\delta\rho}{\rho}\frac{\delta V}{V}\right]
        \\ \vspace{0.2cm}
        \delta\rho_2 \!\!&=&\!\! \displaystyle
        \rho_2 - \rho = \frac{\rho V}{V+\delta V} - \rho
        \approx \rho\left[ - \frac{\delta V}{V}
        + \left(\frac{\delta V}{V}\right)^2\right]
        \\ \vspace{0.2cm}
        \delta B_1 \!\!&=&\!\! \displaystyle
        B_1 - B = \frac{(B+\delta B)(V+\delta V)\varpi}
        {V(\varpi+\delta\varpi)} - B
        \\ \vspace{0.2cm} \!\!&\approx&\!\! \displaystyle
        B\left[\frac{\delta B}{B} + \frac{\delta V}{V} - \frac{\delta\varpi}{\varpi}
        + \frac{\delta B}{B}\frac{\delta V}{V}
        - \frac{\delta\varpi}{\varpi}\frac{\delta B}{B}
        - \frac{\delta\varpi}{\varpi}\frac{\delta V}{V}
        + \left(\frac{\delta\varpi}{\varpi}\right)^2\right]
        \\
        \delta B_2 \!\!&=&\!\! \displaystyle
        B_2 - B = \frac{BV(\varpi+\delta\varpi)}{(V+\delta V)\varpi} - B
        \approx B\left[\frac{\delta\varpi}{\varpi} - \frac{\delta V}{V}
        - \frac{\delta\varpi}{\varpi}\frac{\delta V}{V}
        + \left(\frac{\delta V}{V}\right)^2\right]
        \end{array}
        \enq
Using these and equations (\ref{deltaEp}) and (\ref{deltap}) we can write the energy
change as,
        \beq
        \frac{\Delta E}{V} \approx {\cal M}_0 \left(\frac{\delta V}{V}\right)^2
        + {\cal M}_1\frac{\delta V}{V} + {\cal M}_2 \ ,
        \enq
where,
        \beq
        \begin{array}{l}\vspace{0.2cm}\displaystyle
        {\cal M}_0 \approx
        \gamma p + B^2\frac{\pa^2 F}{\pa B^2}
        + 2\rho B\frac{\pa^2 F}{\pa\rho\pa B} + \rho^2\frac{\pa^2 F}{\pa \rho^2}
        \\ \vspace{0.2cm}\displaystyle
        {\cal M}_1 \approx - 2\rho\delta\phi
        - 2\left(B\frac{\pa F}{\pa B} + B^2\frac{\pa^2 F}{\pa B^2}
        + \rho B\frac{\pa^2 F}{\pa\rho\pa B}\right)\frac{\delta\varpi}{\varpi}
        \\ \displaystyle
        {\cal M}_2 \approx - \delta\rho\delta\phi
        - \left(\frac{\pa F}{\pa B}\delta B
        - B\frac{\pa^2 F}{\pa B^2}\delta B
        - B\frac{\pa^2 F}{\pa\rho\pa B}\delta\rho\right)\frac{\delta\varpi}{\varpi}
        + \left(B\frac{\pa F}{\pa B} + B^2\frac{\pa^2 F}{\pa B^2}\right)
        \left(\frac{\delta\varpi}{\varpi}\right)^2
        \end{array}
        \label{auxM}
        \enq
Since ${\cal M}_0 > 0$ for the cases of interest, the change in energy can be minimized
with respect to $\delta V/V$. The minimum of the energy becomes,
        \beq
        \frac{\Delta E}{V} \approx {\cal M}_2 - \frac{{\cal M}_1^2}{4{\cal M}_0} \ .
        \enq
The small quantities need to be expanded only to first order,
        \beq
        \begin{array}{l}\vspace{0.2cm}\displaystyle
        \delta\varpi = \delta (r\sin\theta) = \delta r\sin\theta +
        r\delta\theta\cos\theta
        \\ \displaystyle
        \delta\rho = \delta r\pa_r\rho + \delta\theta\pa_\theta\rho
        \end{array}
        \enq
and similarly for $B$ and $\phi$. The energy can then be written as,
        \beq
        \frac{\Delta E}{V} \approx a_0 (\delta r)^2 + b_0 r\delta r\delta\theta
        + c_0 r^2 (\delta\theta)^2 \ .
        \enq
$a_0$, $b_0$ and $c_0$ are the same as in equation (\ref{stabilitycoefm0}) and the
conditions for stability are the same as in equation (\ref{con}). In fact, the same
conclusion could have been drawn by comparing equation (\ref{auxM}) to (\ref{auxK}).
Thus, we have shown that the $m = 0$ stability conditions are the same as the conditions
for stability under the interchange of magnetic flux tubes. In other words, the
interchange is the worst instability for $m = 0$.

\subsection{The Muzikar--Pethick--Roberts (MPR) Instability}
In this section we will derive the criteria for the instability discussed by Muzikar \&
Pethick (1981) and Roberts (1981). Using equation (\ref{typeIIfree}) we can write the
magnetic stress tensor as (equation \ref{typeIIstress}),
        \beq
        \sigma_{ij} = \left(F - \rho F_{,\rho}
        - BF_{,B}\right)\delta_{ij}
        + BF_{,B}\hat B_i\hat B_j \ .
        \label{MPRstress}
        \enq
Consider perturbations around a state of uniform density $\rho$ and uniform magnetic
field $\m{B} = B\hz$. The Lagrangian displacement associated with these perturbations is,
        \beq
        \m\xi(\m{r},t) = \m\xi\exp(i\m{k}\cdot\m{r} - i\omega t) \ .
        \enq
In this case, we have,
        \beq
        \begin{array}{l}\vspace{0.2cm}\displaystyle
        \delta\m{B} = \m\nabla\times(\m\xi\times\m{B})
        = i B(k_z\m\xi - \m{k}\cdot\m\xi\hz)
        \\ \vspace{0.2cm}\displaystyle
        \delta B = \hB\cdot\delta\m{B} = - i B\m{k}_\perp\cdot\m\xi_\perp
        \\ \vspace{0.2cm}\displaystyle
        \delta \hB = B^{-1}(\delta\m{B} - \delta B\hB) = i k_z\m\xi_\perp
        \\ \displaystyle
        \delta\rho = - \m\nabla\cdot(\rho\m\xi) = - i \rho\m{k}\cdot\m\xi
        \end{array}
        \label{MPRaux}
        \enq
Here $\perp$ means perpendicular to $\hz$.

The magnetic force density is, from equation (\ref{MPRstress}),
        \beq
        \m{f}_m = \m\nabla\cdot\m\sigma = - \left(\rho F_{,\rho\rho}
        + BF_{,\rho B}\right)\m\nabla\rho
        - \left(\rho F_{,\rho B}
        + BF_{,BB}\right)\m\nabla B
        + \m{B}\cdot\m\nabla(F_{,B}\hB) \ .
        \enq
Since the background quantities are constant the perturbation in the magnetic force
becomes,
        \beq
        \begin{array}{rcl}\vspace{0.2cm}
        \delta\m{f}_m &=& \displaystyle - \left(\rho^2F_{,\rho\rho}
        + \rho BF_{,\rho B}\right)\m{k}(\m{k}\cdot\m\xi)
        - \left(\rho BF_{,\rho B}
        + B^2F_{,BB}\right)\m{k}(\m{k}_\perp\cdot\m\xi_\perp)
        \\ && \displaystyle
        + \hz k_z\left[\rho BF_{,\rho B}(\m{k}\cdot\m\xi)
        + B^2F_{,BB}(\m{k}_\perp\cdot\m\xi_\perp)\right]
        - BF_{,B}k_z^2\m\xi_\perp
        \end{array}
        \enq

Since the background state is symmetric with respect to $\hz$ we can choose $\m{k} = \hz
k_z + \hx k_x$. With this choice equation (\ref{MPRaux}) becomes,
        \beq
        \begin{array}{l}\vspace{0.2cm}\displaystyle
        \delta\m{B} = i B(k_z\xi_x\hx + k_z\xi_y\hy - k_x\xi_x\hz)
        \\ \vspace{0.2cm}\displaystyle
        \delta B = - i B k_x\xi_x
        \\ \vspace{0.2cm}\displaystyle
        \delta \hB = i k_z(\xi_x\hx + \xi_y\hy)
        \\ \displaystyle
        \delta\rho = - i \rho(k_x\xi_x + k_z\xi_z)
        \end{array}
        \enq
The components of the magnetic force become,
        \beq
        \begin{array}{l}\vspace{0.2cm}\displaystyle
        (\delta f_m)_x = - \xi_x\left[k_x^2\left(\rho^2F_{,\rho\rho}
        + 2\rho BF_{,\rho B} + B^2F_{,BB}\right)
        + k_z^2BF_{,B}\right]
        - \xi_zk_xk_z\left(\rho^2F_{,\rho\rho}
        + \rho BF_{,\rho B}\right)
        \\ \vspace{0.2cm}\displaystyle
        (\delta f_m)_y = - \xi_yk_z^2BF_{,B}
        \\ \displaystyle
        (\delta f_m)_z = - \xi_xk_xk_z\left(\rho^2F_{,\rho\rho}
        + \rho BF_{,\rho B}\right)
        - \xi_zk_z^2\rho^2F_{,\rho\rho}
        \end{array}
        \label{MPRfm}
        \enq
In addition, there is a pressure restoring force, $\delta\m{f}_p = - \m\nabla\delta p = -
\gamma p\m{k}(\m{k}\cdot\m\xi)$, or in components,
        \beq
        \begin{array}{l}\vspace{0.2cm}\displaystyle
        (\delta f_p)_x = - \gamma p (k_x^2\xi_x + k_xk_z\xi_z)
        \\ \vspace{0.2cm}\displaystyle
        (\delta f_p)_y = 0
        \\ \displaystyle
        (\delta f_p)_z = - \gamma p (k_xk_z\xi_x + k_z^2\xi_z)
        \end{array}
        \label{MPRfp}
        \enq
We will neglect gravitational forces, so that the equations for the perturbations become,
        \beq
        - \rho\omega^2\m\xi = \delta\m{f}_p + \delta\m{f}_m \ .
        \enq
From equations (\ref{MPRfm}) and (\ref{MPRfp}) it follows that the equation for $\xi_y$
completely decouples from the equations for $\xi_x$ and $\xi_z$,
        \beq
        \rho\omega^2\xi_y = k_z^2BF_{,B}\xi_y \ .
        \enq
This implies that one pair of modes has $\xi_x = \xi_z = 0$ and $\xi_y \ne 0$ with
$\omega^2 = k_z^2BF_{,B}/\rho$. These modes are the generalization of the Alfv\'{e}n
modes. The remaining modes are given through,
        \beq
        \begin{array}{rcl}\vspace{0.2cm}
        \rho\omega^2\xi_x &=& \displaystyle
        \xi_x\left[k_x^2\left(\gamma p + \rho^2F_{,\rho\rho}
        + 2\rho BF_{,\rho B} + B^2F_{,BB}\right)
        + k_z^2BF_{,B}\right]
        + \xi_zk_xk_z\left(\gamma p + \rho^2F_{,\rho\rho}
        + \rho BF_{,\rho B}\right)
        \\
        \rho\omega^2\xi_z &=& \displaystyle
        \xi_xk_xk_z\left(\gamma p + \rho^2F_{,\rho\rho}
        + \rho BF_{,\rho B}\right)
        + \xi_zk_z^2\left(\gamma p + \rho^2F_{,\rho\rho}\right)
        \end{array}
        \enq
From these two equations we get the characteristic equation for the modes, after some
rearrangement,
        \beq
        \rho^2\omega^4 - \rho\omega^2{\cal E}_0 + {\cal E}_1 = 0 \ ,
        \label{MPRmode}
        \enq
where, defining $k^2 = k_x^2 + k_z^2$,
        \beq
        \begin{array}{l}\vspace{0.2cm}\displaystyle
        {\cal E}_0 = k^2\gamma p + k_x^2\left(\rho^2F_{,\rho\rho}
        + 2\rho BF_{,\rho B} + B^2F_{,BB}\right)
        + k_z^2\left(BF_{,B} + \rho^2F_{,\rho\rho}\right)
        \\ \displaystyle
        {\cal E}_1 = k_x^2k_z^2\left(\gamma p B^2F_{,BB} + \rho^2B^2F_{,\rho\rho}F_{,BB}
        - \rho^2B^2F_{,\rho B}^2\right)
        + k_z^4BF_{,B}\left(\gamma p + \rho^2F_{,\rho\rho}\right)
        \end{array}
        \enq
In the absence of magnetic fields, we have, defining $\gamma p = \rho c_s^2$,
        \beq
        \rho^2\omega^4 - \rho^2 \omega^2 k^2 c_s^2 = 0 \ ,
        \enq
which has two roots: $\omega^2 = 0$ and $\omega^2 = k^2c_s^2$. The latter corresponds to
sound waves. In the cases of interest, the magnetic terms will be much smaller in
comparison to the pressure terms, so that one of the roots will have $\omega^2 \approx
k^2c_s^2$ and therefore will be definitely positive. Since ${\cal E}_1$ is the product of
the two roots, the condition for stability is ${\cal E}_1 > 0$, which for $k_z \ne 0$
becomes,
        \beq
        k_x^2\left(\gamma p B^2F_{,BB} + \rho^2B^2F_{,\rho\rho}F_{,BB}
        - \rho^2B^2F_{,\rho B}^2\right)
        + k_z^2BF_{,B}\left(\gamma p + \rho^2F_{,\rho\rho}\right) > 0 \ .
        \enq
For sufficiently large $k_x$, or more precisely when $k_x^2 B F_{,BB} \gg k_z^2F_{,B}$,
this reduces to,
        \beq
        F_{,BB} > \frac{\rho^2 F_{,\rho B}^2}{\gamma p + \rho^2 F_{,\rho\rho}}
        \approx \frac{\rho^2 F_{,\rho B}^2}{\gamma p} \ .
        \label{MPRcon}
        \enq
This is exactly the same condition for stability as in equation (\ref{MPRL0}). When
pressure dominates, it is also of the same form as the condition given by Roberts (1981).
From equation (\ref{MPRmode}) it follows that the potentially unstable modes are given
through,
        \beq
        \rho\omega^2 \approx \frac{{\cal E}_1}{k^2\gamma p}
        \approx \frac{k_x^2k_z^2}{k^2}
        \left(B^2F_{,BB} - \frac{\rho^2B^2F_{,\rho B}^2}{\gamma p}\right)
        + \frac{k_z^4}{k^2}BF_{,B} \ .
        \label{MPRmode2}
        \enq

The magnetic free energy in the strongly type II case ($H \gg B$) can be written as
(Tinkham 1975; Muzikar \& Pethick 1981),
        \beq
        F = \frac{H(\rho)B}{4\pi} + \sqrt{\frac{3}{32\pi^3}}\frac{\Phi_o^2}{\lambda^4}
        \left(\frac{\lambda}{a}\right)^{5/2}\exp\left(-\frac{a}{\lambda}\right) \ .
        \label{MPRfree}
        \enq
Here $\Phi_o = hc/2e$ is the flux quantum ($n_\Phi = B/\Phi_o$ is the flux line density
per unit area), $\lambda = (m_pc^2/4\pi n_pe^2)^{1/2}$ is the London penetration depth,
$n_p$ is the number density of protons, and $a$ is the distance between flux lines in a
triangular lattice,
        \beq
        a = \left(\frac{4}{3}\right)^{1/4}\left(\frac{\Phi_o}{B}\right)^{1/2} \ .
        \label{MPRro}
        \enq
The magnetic field strength in this case is (Tinkham 1975; Easson \& Pethick 1977),
        \beq
        H\simeq H_{\rm c1} = \frac{\Phi_o\ln(\lambda/\xi)}{4\pi\lambda^2} \ ,
        \label{MPRHc1}
        \enq
where $\xi = \hbar^2k_F/\pi m_p\Delta$ is the coherence length, $\xi \ll \lambda$; $k_F =
(3\pi^2n_p)^{1/3}$ is the Fermi wave number of protons, and $\Delta$ is the
superconducting energy gap. The first term in equation (\ref{MPRfree}) is the energy of
an isolated flux line, and the second term arises due to the interaction between flux
lines. Note that only $a$ depends on $B$ and only the interaction term contributes to
$F_{,BB}$. Also note that $\lambda^2 \propto 1/\rho$ when the proton number density is
proportional to the baryon number density, as suggested by Baym et al. (1971). Defining a
new variable by $u = a/\lambda$ we have (equation \ref{MPRfree}),
        \beq
        F = \frac{H(\rho)B}{4\pi} + E(\rho) u^{-5/2} e^{- u} \mtext{where}
        E(\rho) = \sqrt{\frac{3}{32\pi^3}}\frac{\Phi_o^2}{\lambda^4} \ .
        \label{MPRfree2}
        \enq
Then, introducing an auxiliary function $f(u)$,
        \beq
        B^2F_{,BB} = \frac{E(\rho)}{4}\left(u^{-1/2} + 2u^{-3/2} +
        \frac{5}{4}u^{-5/2}\right)e^{-u} = E(\rho)f(u) \ .
        \label{MPRfree3}
        \enq
Only the first term needs to be retained when $u\gg 1$, i.e. when the spacing between
flux lines is large compared to the penetration depth. In the same limit, we can also
approximate,
        \beq
        \rho BF_{,\rho B} \approx \frac{\rho B}{4\pi}\frac{dH}{d\rho} = \frac{\sigma HB}{4\pi}
        \mtext{where} \sigma = \frac{d\ln H}{d\ln\rho} \ .
        \enq
Using these equations, we can write the condition for \emph{instability} as, from
equation (\ref{MPRcon}),
        \beq
        u^4f(u) < \sqrt{\frac{2}{27\pi}} \frac{\sigma^2H^2}{\gamma p} \ .
        \label{MPRcon2}
        \enq
Note than when $\sigma = 0$, i.e. when $H$ is independent of $\rho$, there is no
instability. Thus, it does not arise in a normal medium. Moreover, $\sigma > 0$ is not
required in order to have an instability, contrary to the conclusions of Muzikar \&
Pethick (1981).

We take the magnetic field strength to be $H\sim 10^{15}\,{\rm G}$, and the typical
density in the superconducting region to be $\rho \sim 3\times 10^{14}\,\rm g/cm^{3}$,
which corresponds to a pressure $p\sim 4\times 10^{33}\,{\rm erg/cm^3}$, for a $\gamma =
2$ polytrope and a radius $R_\star \approx 10 \, \rm km$. We also take $\sigma = 1$. From
equation (\ref{MPRcon2}) it follows that instabilities arise for $u>u_o$ where $u_o
\simeq 20$. Using equation (\ref{MPRro}) and the definitions of $\lambda$ and $\Phi_o$,
we can find the largest magnetic induction which is unstable,
        \beq
        B <
        \frac{4\pi h e n_p}{\sqrt{3} m_p c u_o^2} =
        1.15 \times 10^{13}
        \left(\frac{\vphantom{Y}n_p}{0.01 \, \rm fm^{-3}}\right)
        \left(\frac{\vphantom{Y}u_o}{20}\right)^{-2}\rm G \ ,
        \enq
The proton number density $n_p$ is a function of the baryon number density, and for $n_b
\sim 0.2 \, \rm fm^{-3}$, we have $n_p \sim 0.01 \, \rm fm^{-3}$ (Elgar{\o}y et al. 1996;
Zuo et al. 2004).

For toroidal fields $\hz$ is along the $\hphi$ direction, so that for modes we have
$\exp(ik_zz) = \exp(im\phi)$. We can take $k_z \sim m/R_\star$ for a star of radius
$R_\star$. The condition given in equation (\ref{MPRcon}) can lead to instabilities when
the perpendicular wave vector $k_x$ is sufficiently larger than $k_z$. Using equations
(\ref{MPRfree2}) and (\ref{MPRfree3}), we get,
        \beq
        \frac{k_z^2}{k_x^2} \ll \frac{BF_{,BB}}{F_{,B}} \approx
        \frac{4\pi E(\rho)f(u)}{H(\rho)B}
        = \frac{3\sqrt{2\pi} \, u^2f(u)}{\ln(\lambda/\xi)}
        \lesssim \frac{\sigma^2HB}{4\pi\gamma p} \ ,
        \enq
where the last inequality follows from the condition for instability (equation
\ref{MPRcon2}). The length scale of the instabilities is small compared to the size of
the star; for a $\gamma = 2$ polytrope,
        \beq
        L_x = k_x^{-1} \ll \frac{R_\star}{m}\sqrt{\frac{\sigma^2HB}{4\pi\gamma p}}
        \approx 3.1 \times 10^2
        \frac{\displaystyle \sqrt{\sigma^2 H_{15}B_{12}}}{m\rho_{14}} \, \rm cm \, .
        \enq
Here $H_{15} = H/10^{15}\, \rm G$, $B_{12} = B/10^{12}\, \rm G$, and $\rho_{14} =
\rho/10^{14}\, \rm g/cm^3$. From equation (\ref{MPRmode2}) we can estimate the growth
rate of the instability, using $\gamma p = \rho c_s^2$,
        \beq
        \tilde\omega = \sqrt{-\omega^2} \sim \left|\frac{k_z BF_{,\rho B}}{c_s}\right|
        \sim \frac{m|\sigma|HB}{4\pi\rho c_s R_\star} \ .
        \enq
The corresponding growth timescale is,
        \beq
        \frac{1}{\tilde\omega} \approx 3.7 \times 10^3
        \frac{\rho_{14}^{3/2}R_6^2}{m |\sigma| H_{15}B_{12}} \, \rm s \, .
        \enq
Here $R_6 = R_\star/10^6 \, \rm cm$. Note that $m = 0$ is stable. The unstable modes will
be dissipated if the kinematic viscosity of the fluid is,
        \beq
        \eta > \frac{\tilde\omega}{k_x^2} \approx 26 \,
        \frac{H_{15}^2B_{12}^2}{m\rho_{14}^{7/2}R_6^2} \, \rm cm^2/s \, .
        \enq
This value is well below the estimated values of the viscosity in a neutron star, which
are typically in the range $\eta \sim 10^{4-5}\,\rm cm^2/s$ (for a review see Andersson,
Comer \& Glampedakis 2005).

Note that similar results will hold for poloidal fields, except that in this case $k_z
\gtrsim 1/R_\star$ will depend on both the number of radial nodes and the angular
momentum quantum number of the mode. Simple linear analysis along the lines outlined by
Hide (1971) reveals that the growth rate of the MPR mode will not be strongly affected by
buoyancy, but the condition for stability will be modified. Moreover, due to the local
nature of the mode, it is likely to be unaffected by rotation.

\section{Nearly Toroidal Fields}\label{sectionpoloidal}
In normal conducting stars, the presence of poloidal components in addition to toroidal
components may help stabilize the magnetic fields (Tayler 1973; Wright 1973), which has
also been confirmed by recent numerical simulations (Braithwaite \& Nordlund 2006).
Moreover, pulsar observations reveal the presence of a dipole-like field in the neutron
star magnetosphere, implying that a poloidal component of the magnetic field must exist.
The treatment of fully poloidal fields is considerably more complicated and will be
discussed in a subsequent paper. The complication arises as a result of the fact that in
the poloidal case the direction of the magnetic field is not known, and must be computed
numerically (Roberts 1981).

In this section, we will consider the case when there is a small poloidal component in
addition to the much larger toroidal field. We will evaluate the constraints on the shape
of the poloidal component that result from the restrictions that the magnetic force per
unit mass be expressible as a gradient of a potential and that $\m\nabla\cdot\m{B} = 0$.
We will then consider the boundary conditions that must also be satisfied. Let us assume
that the direction of the field is given by,
        \beq
        \hn = \hphi + \m\varepsilon \ ,
        \enq
where $\m\varepsilon$ is a poloidal vector and $|\m\varepsilon| \ll 1$. In what follows,
we will retain only the first order terms in $|\m\varepsilon|$.

The form of the magnetic field inside the superconductor is $\m{H} = H(r,\theta)(\hphi +
\m\varepsilon)$ and the current density can be written as the sum of toroidal and
poloidal components, so instead of equation (\ref{toroidal-current}), we now have,
        \beq
        \begin{array}{l}\vspace{0.2cm}\displaystyle
        \m{J} = \m{J}_{\rm tor} + \m{J}_{\rm pol}
        \\ \vspace{0.2cm}\displaystyle
        \frac{4\pi\m{J}_{\rm tor}}{c} = \m\nabla\times H\hphi
        = \frac{\m\nabla(Hr\sin\theta)\times\hphi}{r\sin\theta}
        \\ \displaystyle
        \frac{4\pi\m{J}_{\rm pol}}{c} = \m\nabla\times H\m\varepsilon
        \end{array}
        \label{poloidal-current}
        \enq
Note that $\m{J}_{\rm tor}$ (due to the toroidal magnetic field) is a poloidal field and
$\m{J}_{\rm pol}$ (due to the poloidal magnetic field) is a toroidal field, i.e.
$\m{J}_{\rm tor}\perp\hphi$ and $\m{J}_{\rm pol}\parallel\hphi$. Taking the induction to
be $\m{B} = B(r,\theta)(\hphi + \m\varepsilon)$, the first term in the force density,
given by equation (\ref{typeIIforce}), becomes,
        \beq
        \frac{\m{J}\times\m{B}}{c} = \frac{\m{J}_{\rm tor}\times B\hphi}{c}
        + \frac{\m{J}_{\rm pol}\times B\hphi}{c}
        + \frac{\m{J}_{\rm tor}\times B\m\varepsilon}{c} \ .
        \label{poloidal-force}
        \enq
The first term is due to the toroidal field, and the second and third term are due to the
presence of the small poloidal component. Since $\m{J}_{\rm pol}$ is a toroidal field the
second term vanishes. On the other hand, the third term is a cross-product of two
poloidal fields, and therefore is a toroidal field. However, we require the toroidal
force density to be zero, so it must vanish. This means that
$\m\varepsilon\parallel\m{J}_{\rm tor}$, or equivalently, in terms of an arbitrary
function $\lambda$,
        \beq
        B(r,\theta)\m\varepsilon = \lambda(r,\theta)\m{J}_{\rm tor} \ .
        \label{poloidal-eps}
        \enq
Thus, the force is of the same form as in the purely toroidal case, and in order for it
to be a gradient, the induction $B$ must still be of the form given by equation
(\ref{toroidal-typeII}). We get a condition on the unknown function $\lambda$ from
$\m\nabla\cdot\m{B} = \m{J}_{\rm tor}\cdot\m\nabla\lambda = 0$,
        \beq
        \frac{4\pi\m{J}_{\rm tor}\cdot\m\nabla\lambda}{c} =
        \frac{\hphi\cdot\m\nabla\lambda\times\m\nabla(Hr\sin\theta)}{r\sin\theta}
        = 0 \ .
        \enq
This equation is satisfied by functions of the form,
        \beq
        \lambda(r,\theta) = \lambda(Hr\sin\theta) \ .
        \enq
Thus, the poloidal vector $\m\varepsilon$ is given by equation (\ref{poloidal-eps}),
using equation (\ref{toroidal-typeII}) for $B$ and equation (\ref{poloidal-current}) for
$\m{J}_{\rm tor}$,
        \beq
        \m\varepsilon = \frac{\lambda(Hr\sin\theta)\m{J}_{\rm tor}}
        {4\pi\rho r\sin\theta f(Hr\sin\theta)} = \frac{H^2}{\rho}
        \m\nabla\tilde\lambda(Hr\sin\theta)\times\hphi \ .
        \label{epsilonvectypeII}
        \enq
In a normal conductor, we have, setting $H = B$ and using equation
(\ref{toroidal-normal}) for $B$,
        \beq
        \m\varepsilon = \frac{\mu(Br\sin\theta)\m{J}_{\rm tor}}{B}
        = \m\nabla\tilde\mu(Br\sin\theta)\times\hphi \ .
        \label{epsilonvecnormal}
        \enq
Here $\mu$ and $\tilde\mu$ are arbitrary functions.

\subsection{Boundary Conditions}
Neglecting second order terms in the small quantity $|\m\varepsilon|$ in the magnetic
stress tensors for the normal and superconducting regions (equations \ref{stressnormal}
and \ref{stresstypeII}), the boundary conditions for the continuity of stress (equation
\ref{stressbc}) become,
        \beq
        - \delta p_s + \sigma_{rr,s} = - \delta p_n + \sigma_{rr,n}
        \mtext{and}
        \sigma_{r\phi,s} = \sigma_{r\phi,n} \ .
        \enq
The $rr$ components of the magnetic stress tensors are the same as in the purely toroidal
case (equation \ref{toroidal-stress}), so the first equation is the same as before
(equation \ref{stressbctor}). However, we now have the second equation, which explicitly
gives, using equations (\ref{stressnormal}) and (\ref{stresstypeII}) for the stress
tensors,
        \beq
        (\hphi\cdot\m{H})(\hr\cdot\m{B}_s) = (\hphi\cdot\m{B}_n)(\hr\cdot\m{B}_n) \ .
        \label{stressbc2}
        \enq
We also have the additional boundary condition on the continuity of the normal component
of the poloidal magnetic induction, which follows from Maxwell's equations,
        \beq
        \hr\cdot\m{B}_s = \hr\cdot\m{B}_n \ .
        \label{maxwell2}
        \enq
The last two equations imply that we must have,
        \beq
        \hphi\cdot\m{H} = \hphi\cdot\m{B}_n \mtext{i.e.} H = B_n \ .
        \enq
This is equivalent to the requirement for the continuity of the $\hphi$ component of the
magnetic field in the absence of surface currents (equation \ref{maxwell}). However, as
was previously discussed, this is inconsistent with our assumption that $H$ is a function
of radius up to the boundaries of the superconductor. This assumption now requires the
presence of a discontinuity in the $\hphi$ component of the magnetic force, although the
forces within the superconducting and normal regions have no such components. This is an
artifact of the incomplete description of the transition boundary, which we have treated
as discontinuous. A more realistic treatment should impose zero toroidal force
everywhere.

Incidentally, note that we cannot simply assume that the radial components of the
poloidal vectors vanish at the boundary, which would also satisfy the above equations
(equations \ref{stressbc2} and \ref{maxwell2}). This would imply that the functions
$\tilde\lambda$ and $\tilde\mu$ in equations (\ref{epsilonvectypeII}) and
(\ref{epsilonvecnormal}) are constants, which in turn would cause the poloidal vectors to
vanish everywhere within the normal and superconducting regions.

\section{Conclusion}
Our main goal in this paper has been to compute the distortion of a neutron star due to a
toroidal magnetic field in its interior, assuming that the star is either partly or
entirely a type II superconductor. Previous authors have estimated the order of magnitude
of this distortion (Jones 1975; Easson \& Pethick 1977; Cutler 2002), finding that it is
enhanced by a factor $H/B$ for given magnetic induction $B$ and magnetic field $H$
compared with the normal case (where $H = B$). In the strongly type II regime, $H \sim
10^{15}\,\rm G$, so that $H/B \sim 10^3/B_{12}$ (Jones 1975; Easson \& Pethick 1977).
Such large enhancements could result in magnetic distortions $\epsilon \sim
10^{-9}-10^{-8}$, which are large enough to be important for neutron star precession
(Wasserman 2003) and possibly for gravitational radiation emission (Cutler 2002). These
earlier works did not compute the structure of the magnetic field in detail.

Here, we have paid closer attention to the requirements of hydrostatic balance and
stability. The assumption of a barotropic equation of state, $p = p(\rho)$, which ought
to apply to a cold neutron star, severely constrains the structure of the toroidal field.
Similar restrictions have been known for a long time for normal conductors (e.g.
Prendergast 1956; Monaghan 1965). The restrictions arise because the magnetic
acceleration must be a total gradient in hydrostatic balance. Under the assumption that
the magnetic free energy $F$ is a function of (matter or baryon) density $\rho$ and
magnetic induction $B$, we find that, for toroidal fields, we must require (equation
\ref{toroidal-typeII}),
        \beq
        B(r,\theta) \propto \rho r\sin\theta f(Hr\sin\theta) \ ,
        \label{toroidal-typeII2}
        \enq
where $f$ is an arbitrary function. Given this function, and $F(\rho,B)$, we can compute
$H(\rho,B) = 4\pi {\pa F/\pa B}$ (equation \ref{typeIIfree}). Equation
(\ref{toroidal-typeII2}) is then an implicit equation that can be used to find
$B(r,\theta)$ (assuming axisymmetry). Similar constraints can be derived for poloidal
magnetic fields, but are more complicated since the field direction must be solved for
(e.g. Roberts 1981 for superconducting, uniform density stars; we will consider
superconducting, barotropic stars in a future paper).

Our calculations have concentrated on neutron stars with a strongly type II regime where
$H$ is independent of $B$; our models allow for as many as two normal regimes interior or
exterior to the superconductor. The main result of these calculations is equation
(\ref{epsilon}) for the magnetic distortion,
        \beq
        \epsilon = 0.945\times 10^{-9} \left(\frac{\phi_2(R_\star)}{\Psi_o}\right)
        \left(\frac{H_c}{10^{15} \, \rm G}\right)
        \left(\frac{B_o}{10^{12} \, \rm G}\right)
        \left(\frac{R_\star}{10 \, \rm km}\right)^4
        \left(\frac{M_\star}{1.4M_\odot}\right)^{-2} \ ,
        \enq
with $\phi_2(R_\star)/\Psi_o \approx - 2$ in all cases, as is summarized in table
\ref{tablephi}. These results were computed for an equation of state $p = \kappa \rho^2$
and $H \propto \rho$ (e.g. Easson \& Pethick 1977). Calculations can be done in a similar
way for other $p(\rho)$ and $H(\rho,B)$.

Although we have separated the star into strongly type II and normal sectors for
computing the deformations due to a toroidal field, we have noted that this assumption,
while mathematically well defined, leads to sudden jumps in density and magnetic
induction at the boundaries of the superconductor. In effect, we have assumed that the
magnetic free energy changes discontinuously from $F = H(\rho)B/4\pi$ in the type II
superconductor to $F = B^2/8\pi$ in the normal conductor. However, our formalism can be
applied more generally to $F(\rho,B)$ that varies smoothly from type II to normal,
probably with intermediate domains of type I superconductivity. Such models ought to be
free of discontinuities in $\rho$ and $B$, but will still have rapid variations in
radially thin domains. In particular, we expect magnetic stresses to be approximately
continuous across boundaries, so the magnetic induction $B_n$ in the normal regions will
be larger than the induction $B_s$ in the superconductor, $B_n \sim (HB_s)^{1/2} \gg
B_s$. Strong toroidal fields $B_n \sim 10^{13.5}\, \rm G$ (corresponding to $H \sim
10^{15}\, \rm G$ and $B_s \sim 10^{12}\, \rm G$) are needed for large distortions;
toroidal fields $B_n \sim 10^{12}\, \rm G$ imply $B_s \sim 10^{9}\, \rm G$ and therefore
will lead to $\epsilon \sim 10^{-12}$. We have postponed considering models with
realistic $F(\rho,B)$, which would be more intricate mathematically, to later work.

A toroidal field can be produced as a result of the winding up of the magnetic field
early in the history of a neutron star (Thompson \& Duncan 2001). The resulting field
could be stronger than $10^{12}\, \rm G$. When the star has cooled down sufficiently, the
superconducting shell forms. This would produce a large stress within the superconductor
and the star would become dynamically unstable. This, in turn, would lead to a lowering
of the induction inside the superconductor until stability can be restored. In
equilibrium, the stresses within the superconductor and the normal regions will be
comparable. In other words, the amplitude of the magnetic stress may be fixed by the
original amplification of the toroidal field. The superconductor adjusts to the
requirement of approximately continuous stress by lowering $B_s$. In this sense, the
superconductor doesn't really amplify the stress.

Magnetic fields not only need to be in magnetohydrostatic equilibrium, but they must also
be stable with respect to perturbations. We have derived stability criteria from an
energy principle for generic $F(\rho,B)$. This is more general than the treatment of
Roberts (1981), who assumed $H\propto\rho$, and it also includes the normal case treated
previously by Tayler (1973) as the special case $H = B$. In a completely type II
superconducting star with $H \propto \rho$ and $B \propto \sin^n\theta$ (equation
\ref{toroidal-typeIIB}), we find that only $n = 1$ is stable to $m = 0$ (axisymmetric)
perturbations. In fact, as we show in Appendix A, this is true for any magnetic field of
the form $H(\rho,B)$. For $m \ne 0$ all field configurations in a type II star are prone
to the Muzikar--Pethick--Roberts (MPR) instability, found by Muzikar \& Pethick (1981)
and Roberts (1981), when $B \lesssim 10^{13}\,\rm G$. There is also a minimum wave number
for instability, and it is very large: the MPR instability is a small scale instability.
From a linear perturbation analysis around a uniform background, we find that the
instability has a length scale $\sim 10^{-4} R_\star$, where $R_\star$ is the stellar
radius, and a timescale $\sim 10^3 \, \rm s$. This timescale is relatively long compared
to an Alfv\'{e}n crossing time $t_A = R_\star(4\pi\rho/HB)^{1/2} \approx 3.5 \,
R_6(\rho_{15}/H_{15}B_{12})^{1/2} \, \rm s$, but short compared to a typical precession
period of the order of a year. We have also argued that the MPR instability cannot occur
for $m = 0$ in toroidal fields: our linear analysis implies zero growth rate for modes
with wave vectors entirely orthogonal to the unperturbed magnetic field. Because of the
large wave numbers required for the instability, viscous effects, which cannot be studied
via stability analyses from energy principles, could prevent it from occurring
altogether. Our estimate is that a kinematic viscosity of $\sim 10-100\,{\rm cm^2/s}$
would be enough to shut off the instability; this value is smaller than most estimates of
the kinematic viscosity in neutron star matter (Andersson et al. 2005).

We find that normal toroidal fields with $B \propto \sin^{n/2}\theta$ (equation
\ref{toroidal-normalB}) are unstable for $m=1$. Therefore, toroidal fields in a star with
normal and superconducting regions will be unstable. Poloidal fields may help stabilize
the stellar magnetic field, as has been found for normal conductors (e.g. Tayler 1973;
Wright 1973; Braithwaite \& Nordlund 2006). Moreover, the emission from radio pulsars
additionally requires exterior, poloidal fields. Consequently, we have also considered
nearly toroidal fields in which the field direction is $\hphi + \m\varepsilon$, where
$\m\varepsilon \perp \hphi$ and $|\m\varepsilon| \ll 1$. Here, too, the form of
$\m\varepsilon$ is not completely arbitrary: to maintain hydrostatic balance and
eliminate toroidal forces, we find the requirement (equation \ref{epsilonvectypeII}),
        \beq
        \m\varepsilon = \frac{H^2}{\rho}
        \m\nabla\tilde\lambda(Hr\sin\theta)\times\hphi \ ,
        \label{epsilonvectypeII2}
        \enq
where $\tilde\lambda$ is an arbitrary function. We derived equation
(\ref{epsilonvectypeII2}) for type II regimes, but it holds elsewhere (in particular, in
normal regions). We have seen, though, that when we assume discontinuous transitions in
the magnetic free energy between type II and normal regions, there are discontinuities in
the $r\phi$ component of the magnetic stress tensor, implying a surface toroidal force. A
more complete treatment with continuously varying $F(\rho,B)$ would not have such surface
forces since equation (\ref{epsilonvectypeII2}) would then guarantee vanishing toroidal
forces everywhere.

The results found here can be applied directly to precession of neutron stars. For fluid
stars, Spitzer (1958) argued that precession is inevitable if the magnetic and rotational
axes are misaligned; Mestel \& Takhar (1972) showed that the star precesses about its
magnetic symmetry axis with a period $P_p = P_\star / 3\epsilon_{\rm mag}\cos\chi$ where
$\chi$ is the misalignment angle. For a radio pulsar, there would be no effect on the
arrival times of pulses if the pulsar beam is along the magnetic axis of the star.
Wasserman (2003) showed that crustal distortions with a symmetry axis that is also
misaligned with the magnetic axis would lead to periodically varying timing residuals.
For PSR B1828--11, spindown can enhance the effect considerably, and the data can be
accounted for with $B\sim 10^{12-13}\, \rm G$, $\chi\sim 1 \, \rm rad$, and a modest
permanent crustal distortion $\sim 0.01$ times the magnetic distortion. (Perhaps
fortuitously, this is close to the crustal distortion found by Cutler et al. 2003 for
relaxation near the actual rotation frequency of PSR B1828--11.) The model favors prolate
figures (see also Akg\"{u}n et al. 2006), as would be expected from (predominantly)
toroidal fields. Why the magnetic and spin axes are misaligned remains unexplained.
Moreover, the effects of the slow, time variable fluid motions that would be required in
such a model (e.g. Mestel \& Takhar 1972; Mestel et al. 1981; Nittmann \& Wood 1981) have
yet to be computed.

In this paper, we have not examined the effects of rotation, internal velocity fields,
multi-fluid components, drag and dissipation. These will likely introduce new modes and
will alter the properties of modes of non-rotating stars.

\section*{Acknowledgements}
This research is supported in part by NSF AST-0307273 and 0606710 (Cornell University).
We would like to thank the referee for useful comments on the manuscript.

\appendix

\section{Stability Criteria for a Magnetic Field $H(\rho,B)$}
The coefficients for $m = 0$ for a magnetic free energy $F(\rho,B)$ are given by equation
(\ref{stabilitycoefm0}), where the various quantities are defined in equations
(\ref{auxK}) and (\ref{auxU}). The hydrostatic equilibrium in the absence of magnetic
fields is spherically symmetric, $\pa_r p_o + \rho_o\pa_r\phi_o = 0$. In the presence of
magnetic fields, the equilibrium is given by equation (\ref{euler2var}). Using these
equations, we can rewrite the coefficients as, to lowest order in $F$,
        \beq
        \begin{aligned}
        a_0 & \approx T_0\left(\frac{d\ln\rho}{dr}\right)^2
        + T_1\frac{d\ln\rho}{dr} - U_1 + U_3 \\
        b_0 & \approx T_2\frac{d\ln\rho}{dr} - U_1\cot\theta - U_2 + 2U_3\cot\theta \\
        c_0 & \approx - U_2\cot\theta + U_3\cot^2\theta
        \end{aligned}
        \label{coeff}
        \enq
where,
        \beq
        \begin{aligned}
        T_0 & = B^2F_{,BB} + 2\rho BF_{,\rho B} + \rho^2F_{,\rho\rho} \\
        T_1 & = 2U_0 - \frac{BF_{,B}}{r} - B\pa_r BF_{,BB} - B\pa_r\rho F_{,\rho B}
        - \rho\pa_r BF_{,\rho B} - \rho\pa_r\rho F_{,\rho\rho} \\
        T_2 & = 2U_0\cot\theta - \frac{1}{r}(BF_{,B}\cot\theta + B\pa_\theta BF_{,BB}
        + B\pa_\theta\rho F_{,\rho B} + \rho\pa_\theta BF_{,\rho B}
        + \rho\pa_\theta\rho F_{,\rho\rho})
        \end{aligned}
        \label{auxT}
        \enq

Consider the case of a magnetic field $H(\rho,B)$. In this case, the magnetic free energy
$F(\rho,B)$ is given through $H = 4\pi F_{,B}$ (equation \ref{typeIIfree}). To lowest
order in $F$, the density is a function of radius, $\rho(r)$. Therefore, partial
derivatives of $\rho$ with respect to the angle $\theta$ can be dropped. Then, equation
(\ref{coeff}) can be written equivalently as,
        \beq
        \begin{aligned}
        a_0 & \approx \frac{1}{r^2}\left[Q_1\left(\frac{d\ln\rho}{d\ln r}\right)^2
        + Q_2\frac{d\ln\rho}{d\ln r} + Q_3\right] \\
        b_0 & \approx \frac{\cot\theta}{r^2}\left[Q_4\frac{d\ln\rho}{d\ln r}
        + Q_3 + Q_5\right] \\
        c_0 & \approx \frac{\cot^2\theta}{r^2}Q_5
        \end{aligned}
        \label{coeffalt}
        \enq
where, we define,
        \beq
        \begin{aligned}
        Q_0 & = BF_{,B} + B^2F_{,BB} & \hspace{1.2cm}
        Q_2 & = Q_0 + Q_1\left(1 - \frac{\pa\ln B}{\pa\ln r}\right) & \hspace{1.2cm}
        Q_4 & = Q_0 + Q_1\left(1 - \frac{\pa\ln B}{\pa\ln\sin\theta}\right) \\
        Q_1 & = B^2F_{,BB} + \rho BF_{,\rho B} &
        Q_3 & = Q_0\left(1 - \frac{\pa\ln B}{\pa\ln r}\right) &
        Q_5 & = Q_0\left(1 - \frac{\pa\ln B}{\pa\ln\sin\theta}\right)
        \end{aligned}
        \enq

The magnetic induction is given by equation (\ref{toroidal-typeII}), which we can rewrite
in terms of a new arbitrary function $g$ as,
        \beq
        B(r,\theta) = \frac{\rho g(Hr\sin\theta)}{H} \ .
        \label{formB}
        \enq
Let $\zeta = Hr\sin\theta$ be the argument of the function $g$, and define,
        \beq
        \eta = \frac{d\ln g}{d\ln\zeta} \ , \hspace{0.6cm}
        \xi = \frac{d\ln\rho}{d\ln r} \ , \hspace{0.6cm}
        \sigma_\rho = \frac{\pa\ln H}{\pa\ln\rho} \mtext{and}
        \sigma_B = \frac{\pa\ln H}{\pa\ln B} \ .
        \enq
Then, after some algebra it follows that,
        \beq
        \frac{\pa\ln B}{\pa\ln r} =
        \frac{\xi(1 - \sigma_\rho) + \eta(1 + \xi\sigma_\rho)}
        {1 + \sigma_B(1 - \eta)} \mtext{and}
        \frac{\pa\ln B}{\pa\ln\sin\theta} = \frac{\eta}{1 + \sigma_B(1 - \eta)} \ .
        \label{derB}
        \enq
Using $H = 4\pi F_{,B}$, we also get,
        \beq
        Q_0 = \frac{HB}{4\pi}(1 + \sigma_B) \mtext{and}
        Q_1 = \frac{HB}{4\pi}(\sigma_\rho + \sigma_B) \ .
        \enq
Then, the coefficients become (from equation \ref{coeffalt}),
        \beq
        \begin{aligned}
        a_0 & \approx \frac{HB}{4\pi r^2}
        \frac{[1 + \sigma_B + \xi(\sigma_\rho + \sigma_B)]^2(1 - \eta)}
        {1 + \sigma_B(1 - \eta)} \\
        b_0 & \approx \frac{HB\cot\theta}{2\pi r^2}
        \frac{[1 + \sigma_B + \xi(\sigma_\rho + \sigma_B)](1 + \sigma_B)(1 - \eta)}
        {1 + \sigma_B(1 - \eta)} \\
        c_0 & \approx \frac{HB\cot^2\theta}{4\pi r^2}
        \frac{(1 + \sigma_B)^2(1 - \eta)}{1 + \sigma_B(1 - \eta)}
        \end{aligned}
        \enq
Since $b_0\!^2 = 4a_0c_0$, one of the stability conditions is immediately marginally
satisfied. The other two conditions give,
        \beq
        \frac{1 - \eta}{1 + \sigma_B(1 - \eta)} > 0 \ .
        \label{condition}
        \enq
Using equation (\ref{derB}) we can rewrite this condition as, for $1 + \sigma_B > 0$,
        \beq
        \frac{\pa\ln B}{\pa\ln\sin\theta} < 1 \ .
        \enq
Thus, the magnetic fields are marginally stable for $B \propto \sin\theta$.

For a strongly type II superconducting star $H = H(\rho)$, so that $\sigma_B = 0$, and
equation (\ref{condition}) reduces to $\eta < 1$. For a normal conducting star $H = B$,
so that $\sigma_B = 1$, and we get $(1 - \eta)/(2 - \eta) > 0$. This condition can be
expressed in an alternative way by noting that equation (\ref{formB}) for a normal
conductor is $B = \rho g(Br\sin\theta)/B$. Thus, $B$ is given as a function of itself.
This equation can be rewritten as $B = h(\rho r^2\sin^2\theta)/r\sin\theta$, and the
magnetic free energy is given by $F = B^2/8\pi = \rho f(\rho r^2\sin^2\theta)$, where $h$
and $f$ are arbitrary functions. From here and from equation (\ref{derB}) it follows
that, defining $\varpi = \rho r^2\sin^2\theta$,
        \beq
        \frac{\pa\ln B}{\pa\ln\sin\theta} = \frac{d\ln f}{d\ln\varpi}
        = \frac{\eta}{2 - \eta} \ .
        \enq
The same result is obtained by considering the derivative of $B$ with respect to $r$,
though it involves more algebra. Thus, the stability condition for the normal conducting
case is better expressed as,
        \beq
        \frac{d\ln f}{d\ln\varpi} < 1 \ .
        \enq
For a normal conducting star, the field is marginally stable for $f \propto \varpi$, i.e.
$B \propto \rho r\sin\theta$, as noted in \S\ref{sectioncritnormal}. Similarly, for a
strongly type II superconducting star, the field is marginally stable for $g \propto
\zeta$, i.e. $B \propto \rho r\sin\theta$, as noted in \S\ref{sectioncrittypeII}.


\begin{thebibliography}{}
\bibitem[Akg\"{u}n 2007]{akgunref}
        Akg\"{u}n T., 2007, PhD thesis, Cornell University
\bibitem[Akg\"{u}n et al. 2006]{akgunref2}
        Akg\"{u}n T., Link B., Wasserman I., 2006, MNRAS, 365, 653
\bibitem[Andersson et al. 2005]{anderssonref}
        Andersson N., Comer G. L., Glampedakis K., 2005, Nucl. Phys. A, 763, 212
\bibitem[Baldo \& Schulze 2007]{baldoref}
        Baldo M., Schulze H.--J., 2007, Phys. Rev. C, 75, 025802
\bibitem[Baym \& Pethick 1975]{baymref}
        Baym G., Pethick C. J., 1975, Ann. Rev. Nucl. Sci., 25, 27
\bibitem[Baym \& Pines 1971]{baymref2}
        Baym G., Pines D., 1971, Ann. Phys., 66, 816
\bibitem[Baym et al. 1969]{baymref3}
        Baym G., Pethick C. J., Pines D., 1969, Nature, 224, 673
\bibitem[Baym et al. 1971]{baymref4}
        Baym G., Bethe H. A., Pethick C. J., 1971, Nucl. Phys. A, 175, 225
\bibitem[Bernstein et al. 1958]{bernsteinref}
        Bernstein I. B., Frieman E. A., Kruskal M. D., Kulsrud R. M., 1958, Proc. R. Soc. A, 244, 17
\bibitem[Braithwaite \& Nordlund 2006]{braithwaiteref}
        Braithwaite J., Nordlund {\AA}., 2006, A\&A, 450, 1077
\bibitem[Cordes 1993]{cordesref}
        Cordes J. M., 1993,
        in Phillips J. A., Thorsett S. E., Kulkarni S. R., eds, ASP Conf. Ser. Vol. 36, Planets
        around Pulsars. Astron. Soc. Pac., San Francisco, p. 43
\bibitem[Cutler 2002]{cutlerref}
        Cutler C., 2002, Phys. Rev. D, 66, 084025
\bibitem[Cutler et al. 2003]{cutlerref2}
        Cutler C., Ushomirsky G., Link B., 2003, ApJ, 588, 975
\bibitem[Easson \& Pethick 1977]{eassonref}
        Easson I., Pethick C. J., 1977, Phys. Rev. D, 16, 275
\bibitem[Elgar{\o}y et al. 1996]{elgaroyref}
        Elgar{\o}y {\O}., Engvik L., Hjorth--Jensen M., Osnes E., 1996, Phys. Rev. Lett.,
        77, 1428
\bibitem[Ferri\`{e}re et al. 1999]{ferriereref}
        Ferri\`{e}re K. M., Zimmer C., Blanc M., 1999, J. Geophys. Res., 104, 17335
\bibitem[Ferri\`{e}re et al. 2001]{ferriereref2}
        Ferri\`{e}re K. M., Zimmer C., Blanc M., 2001, J. Geophys. Res., 106, 327
\bibitem[Freidberg 1982]{freidbergref}
        Freidberg J. P., 1982, Rev. Mod. Phys., 54, 801
\bibitem[Friedman \& Schutz 1978]{friedmanref}
        Friedman J. L., Schutz B. F., 1978, ApJ, 221, 937
\bibitem[Glampedakis \& Andersson 2007]{glampedakisref}
        Glampedakis K., Andersson N., 2007, MNRAS, 377, 630
\bibitem[Glampedakis et al. 2007]{glampedakisref2}
        Glampedakis K., Andersson N., Jones D. I., 2007, preprint arXiv:0708.2693
\bibitem[Goossens \& Veugelen 1978]{goossensref}
        Goossens M., Veugelen P., 1978, A\&A, 70, 277
\bibitem[Hide 1971]{hideref}
        Hide R., 1971, QJRAS, 12, 380
\bibitem[Ioka 2001]{iokaref}
        Ioka K., 2001, MNRAS, 327, 639
\bibitem[Jones 1975]{jonesref}
        Jones P. B., 1975, Ap\&SS, 33, 215
\bibitem[Jones 2006]{jonesref2}
        Jones P. B., 2006, MNRAS, 365, 339
\bibitem[Jones \& Andersson 2001]{anderssonref2}
        Jones D. I., Andersson N., 2001, MNRAS, 324, 811
\bibitem[Josephson 1966]{josephsonref}
        Josephson B. D., 1966, Phys. Rev., 152, 1
\bibitem[Link 2003]{linkref}
        Link B., 2003, Phys. Rev. Lett., 91, 101101
\bibitem[Link \& Cutler 2002]{linkref2}
        Link B., Cutler C., 2002, MNRAS, 336, 211
\bibitem[Link \& Epstein 2001]{linkref3}
        Link B., Epstein R. I., 2001, ApJ, 556, 392
\bibitem[Lorenz et al. 1993]{lorenzref}
        Lorenz C. P., Ravenhall D. G., Pethick C. J., 1993, Phys. Rev. Lett., 70, 379
\bibitem[Mestel \& Takhar 1972]{mestelref}
        Mestel L., Takhar H. S., 1972, MNRAS, 156, 419
\bibitem[Mestel et al. 1981]{mestelref2}
        Mestel L., Nittmann J., Wood W. P., Wright G. A. E., 1981, MNRAS, 195, 979
\bibitem[Monaghan 1965]{monaghanref}
        Monaghan J. J., 1965, MNRAS, 131, 105
\bibitem[Muzikar \& Pethick 1981]{muzikarref}
        Muzikar P., Pethick C. J., 1981, Phys. Rev. B, 24, 2533
\bibitem[Nittmann \& Wood 1981]{nittmannref}
        Nittmann J., Wood W. P., 1981, MNRAS, 196, 491
\bibitem[Prendergast 1956]{prendergastref}
        Prendergast K. H., 1956, ApJ, 123, 498
\bibitem[Reisenegger \& Goldreich 1992]{reiseneggerref}
        Reisenegger A., Goldreich P., 1992, ApJ, 395, 240
\bibitem[Roberts 1981]{robertsref}
        Roberts P. H., 1981, Q. J. Mech. Appl. Math., Vol. XXXIV, Pt. 3
\bibitem[Sedrakian et al. 1999]{sedrakianref}
        Sedrakian A. D., Wasserman I., Cordes J. M., 1999, ApJ, 524, 341
\bibitem[Shaham 1977]{shahamref}
        Shaham J., 1977, ApJ, 214, 251.
\bibitem[Shaham 1986]{shahamref2}
        Shaham J., 1986, ApJ, 310, 708.
\bibitem[Spitzer 1958]{spitzerref}
        Spitzer L., 1958, IAU Symp. 6,
        Electromagnetic Phenomena in Cosmical Physics, 169
\bibitem[Stairs et al. 2000]{stairsref}
        Stairs I. H., Lyne A. G., Shemar S. L., 2000, Nature, 406, 484
\bibitem[Stairs et al. 2003]{stairsref2}
        Stairs I. H., Athanasiadis D., Kramer M., Lyne A. G., 2003,
        in Bailes M., Nice D. J., Thorsett S. E., eds, ASP Conf. Ser. Vol. 302, Radio Pulsars.
        Astron. Soc. Pac., San Francisco, p. 249
\bibitem[Tayler 1973]{tayler}
        Tayler R. J., 1973, MNRAS, 161, 365
\bibitem[Thompson \& Duncan 2001]{thompsonref}
        Thompson C., Duncan R. C., 2001, ApJ, 561, 980
\bibitem[Tinkham 1975]{tinkhamref}
        Tinkham M., 1975, Introduction to Superconductivity. McGraw--Hill, New York
\bibitem[Wasserman 2003]{wassermanref}
        Wasserman I., 2003, MNRAS, 341, 1020
\bibitem[Wright 1973]{wrightref}
        Wright G. A. E., 1973, MNRAS, 162, 339
\bibitem[Yakovlev \& Pethick 2004]{yakovlevref}
        Yakovlev D. G., Pethick C. J., 2004, ARA\&A, 42, 169
\bibitem[Zuo et al. 2004]{lombardoref}
        Zuo W., Li Z. H., Lu G. C., Li J. Q., Scheid W., Lombardo U., Schulze H.--J., Shen C.
        W., 2004, Phys. Lett. B, 595, 44
\end{thebibliography}
\end{document}